\pdfoutput=1
\newcommand*{\ATLASLATEXPATH}{}
\documentclass[cernpreprint,texlive=2016,txfonts,UKenglish]{\ATLASLATEXPATH atlasdoc}
 
\usepackage[subfigure,block=none]{\ATLASLATEXPATH atlaspackage}
\usepackage{\ATLASLATEXPATH atlasbiblatex}
 
\usepackage{\ATLASLATEXPATH atlascontribute}
 
\usepackage{\ATLASLATEXPATH atlasphysics}
 
\graphicspath{{logos/}{figures/}}
 
\usepackage{multib-defs}
 
\usepackage{\ATLASLATEXPATH atlasbsm}
 
\usepackage{float}
\usepackage{multirow}
\usepackage{tikz}
\usepackage{amssymb}
\usepackage{lscape}
\usepackage{array}
\usepackage{tabulary}
\newcolumntype{K}[1]{>{\centering\arraybackslash}p{#1}}
\usepackage{amssymb}
\usepackage{pifont}
\newcommand{\cmark}{\ding{51}}

\AtlasTitle{Search for supersymmetry in final states with missing
transverse momentum and multiple $b$-jets in proton--proton collisions
at $\sqrt{s} = 13$ TeV with the ATLAS detector}
 
\author{The ATLAS Collaboration}
 
\AtlasNote{SUSY-2016-10}
\PreprintIdNumber{CERN-EP-2017-182}
 
\AtlasJournalRef{JHEP 06 (2018) 107}
\AtlasDOI{DOI:10.1007/JHEP06(2018)107}
 
\AtlasAbstract{
A search for supersymmetry involving the pair production of gluinos decaying via third-generation squarks into the lightest neutralino ($\displaystyle\tilde\chi^0_1$) is reported. It uses LHC proton--proton collision data at a centre-of-mass energy $\sqrt{s} = 13$ TeV with an integrated luminosity of 36.1 fb$^{-1}$ collected with the ATLAS detector in 2015 and 2016. The search is performed in events containing large missing transverse momentum and several energetic jets, at least three of which must be identified as originating from $b$-quarks. To increase the sensitivity, the sample is divided into subsamples based on the presence or absence of electrons or muons. No excess is found above the predicted background.  For $\displaystyle\tilde\chi^0_1$ masses below approximately 300 GeV, gluino masses of less than 1.97 (1.92) TeV are excluded at 95\% confidence level in simplified models involving the pair production of gluinos that decay via top (bottom) squarks. An interpretation of the limits in terms of the branching ratios of the gluinos into third-generation squarks is also provided. These results improve upon the exclusion limits obtained with the 3.2 fb$^{-1}$ of data collected in 2015.
}
\hypersetup{pdftitle={ATLAS document},pdfauthor={The ATLAS Collaboration}}
 
\begin{document}
 
\maketitle

\section{Introduction}
\label{sec:intro}
Supersymmetry (SUSY)~\cite{Golfand:1971iw,Volkov:1973ix,Wess:1974tw,Wess:1974jb,Ferrara:1974pu,Salam:1974ig}
is a generalisation of space-time symmetries that predicts new bosonic partners for the fermions and new fermionic
partners for the bosons of the Standard Model (SM).
If $R$-parity is conserved~\cite{Farrar:1978xj}, SUSY particles are produced in pairs and the lightest
supersymmetric particle (LSP) is stable. The scalar partners of the left- and right-handed quarks, the squarks
$\squarkL$ and $\squarkR$, can mix to form two mass eigenstates $\tilde{q}_1$ and $\tilde{q}_2$, ordered by
increasing mass.
SUSY can solve the hierarchy problem~\cite{Sakai:1981gr,Dimopoulos:1981yj,Ibanez:1981yh,Dimopoulos:1981zb} reducing
unnatural tuning in the Higgs sector by orders of magnitude, provided that the superpartners of the top quark have masses not too far above the weak scale.
The large top Yukawa coupling results in significant $\stopL$-$\stopR$ mixing so that the mass eigenstate $\stopone$ is typically lighter than the other squarks \cite{Inoue:1982pi,Ellis:1983ed}.
Because of the  SM weak-isospin symmetry,
the mass of the lightest bottom squark $\sbottomone$ is also
expected to be close to the weak scale. The fermionic partners of the gluons, the gluinos ($\gluino$), are also
motivated by naturalness~\cite{Barbieri:1987fn} to have a mass around the \tev\ scale in order to limit their
contributions to the radiative corrections to the top squark masses. For these reasons, and because the gluinos are
expected to be pair-produced with a high cross-section at the Large Hadron Collider (LHC), the search for gluino
production with decays via top and bottom squarks is highly motivated at the LHC.
 
This paper presents a search for pair-produced gluinos decaying via top or bottom squarks in events
with multiple jets originating from the hadronisation of $b$-quarks ($b$-jets in the following), high missing
transverse momentum of magnitude \met, and potentially additional light-quark jets and/or an isolated charged lepton.\footnote{The term ``lepton'' refers exclusively to an electron or a muon in this paper.}
The dataset consists of 36.1~\ifb\ of
proton--proton ($pp$) collision data collected with the ATLAS detector~\cite{PERF-2007-01} at a centre-of-mass
energy of 13~\tev\ in 2015 and 2016.
Interpretations are provided in the context of several effective simplified models \cite{Alwall:2008ve,Alwall:2008ag,Alves:2011wf} probing various gluino decays into
third-generation squarks and the LSP. The latter is assumed to be the lightest neutralino $\ninoone$, a linear
superposition of the superpartners of the neutral electroweak and Higgs bosons. One model also features the lightest charginos $\chinoonepm$,  which are linear
superpositions of the superpartners of the charged electroweak and Higgs bosons.
The results supersede the ones obtained using 3.2~\ifb\ of data collected in 2015 using the same strategy \cite{Aad:2016eki}.
Pair-produced gluinos with top-squark-mediated decays have also been searched for using events containing pairs of same-sign leptons or three leptons using 13~\tev\ data \cite{Aad:2016tuk,Aaboud:2017dmy}.
The same-sign/three lepton search is comparable in sensitivity to the search presented in this paper only when the masses of the gluino and the LSP are very close to each other.
Similar searches performed using the 13~\tev\ dataset collected in 2015 and 2016 by the CMS experiment have produced results comparable to the ATLAS searches \cite{Sirunyan:2017uyt,Sirunyan:2017fsj,Sirunyan:2017cwe,Sirunyan:2017kqq}.
 
 
\section{SUSY signal models}
\label{sec:susy_sig}
 
Various simplified SUSY models  \cite{Alwall:2008ag,Alves:2011wf} are employed to optimise the event selection and/or interpret the results of the search.
In terms of experimental signature, they all contain at least four $b$-jets originating from either gluino or top quark decays, and two $\ninoone$, which escape the detector unseen, resulting in high $\met$.
 
Gluinos are assumed to be pair-produced and to decay either as  $\gluino \to \sbottomone \bar{b}$ or $\gluino \to \stopone \bar{t}$ (the charge conjugate process is implied throughout this paper). The following
top and bottom squark decays are then considered: $\stopone \to t \ninoone$, $\stopone \to b \chinoonep$ and $\sbottomone \to b \ninoone$.\footnote{The decay $\sbottomone \to t \chinoonem$ is also possible but, following $\gluino \to \sbottomone \bar{b}$, it yields the same final state as  $\gluino \to \stopone^* t \to ( \bar{b} \chinoonem) t$, which is already considered.}
In all cases, the top or bottom squarks are assumed to be off-shell in order to have simplified models with only two parameters: the gluino
and $\ninoone$ masses.\footnote{The analysis sensitivity is found to be mostly independent of the top and bottom squark masses, except when the top squark is very light \cite{3bjetsPaper}.}
All other sparticles are decoupled.
 
Two simplified models are used to optimise the event selection and to interpret the results. In the Gbb (Gtt) model,
illustrated in Figure~\ref{fig:feynman-Gluino-topbottom-offshell}(a) (\ref{fig:feynman-Gluino-topbottom-offshell}(b)), each gluino undergoes an effective three-body
decay $\gluino \to b\bar{b}\ninoone$ ($\gluino \to t\bar{t}\ninoone$) via off-shell bottom (top) squarks, with a branching ratio of 100\%.
The Gbb model is the simplest in terms of particle multiplicity, resulting in the minimal common features of four $b$-jets and two $\ninoone$. In addition to these particles, the Gtt model produces four $W$ bosons originating from the top quark decays: $t \to W b$. The presence of these four $W$ bosons motivates the design of signal regions with a higher jet multiplicity than for Gbb models, and in some cases with at least one isolated electron or muon.

\begin{figure*}[h]
\centering
\subfigure[]{\includegraphics[width=0.35\textwidth]{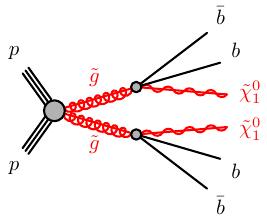}}
\subfigure[]{\includegraphics[width=0.35\textwidth]{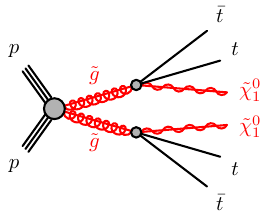}}
\caption{The decay topologies in the (a) Gbb and (b) Gtt simplified models.
}\label{fig:feynman-Gluino-topbottom-offshell}
\end{figure*}
 
This paper includes an interpretation that probes the sensitivity of the search as a function of the gluino branching ratio, in addition to the gluino and $\ninoone$ masses. Similar interpretations have been performed by the CMS collaboration \cite{Khachatryan:2016epu,Sirunyan:2017cwe}.
For that interpretation a third gluino decay is considered:  $\gluino \to t \bar{b} \chinoonem$ (via the off-shell top squark decay $\stopone^* \to \bar{b} \chinoonem$). The $\chinoonem$ is then forced
to decay as $\chinoonepm \to W^{*} \ninoone \to f\bar{f}' \ninoone $ (where $f$ denotes a fermion). To keep the number of model parameters at only two, the mass difference between the $\chinoonepm$ and the $\ninoone$ is
fixed to 2 GeV. Such a small mass-splitting between the $\chinoonepm$ and the $\ninoone$ is typical of models where the $\ninoone$ is dominated by the higgsinos, the superpartners of the neutral Higgs boson. Such models are well motivated by naturalness.
The products of the decay $ W^* \to f\bar{f}'$ are typically too soft to be detected, except for very large mass differences between the gluino and the $\chinoonepm$.
Thus, in this model, the gluino can decay as either $\gluino \to b\bar{b}\ninoone$, $\gluino \to t \bar{b} \chinoonem$ (with $\chinoonem \to f\bar{f}' \ninoone)$ or $\gluino \to t\bar{t}\ninoone$, with the sum of individual branching ratios adding up
to 100\%. This model probes more realistic scenarios where the branching ratio for either $\gluino \to b\bar{b}\ninoone$ or $\gluino \to t\bar{t}\ninoone$ is
not 100\%, and where one, two or three top quarks, and thus on-shell $W$ bosons, are possible in the final state, in between the Gbb (no top quarks) and Gtt (four top quarks) decay topologies.
The decay topologies that are considered in the variable branching ratio model are illustrated in Figure~\ref{fig:feynman-Gluino-topbottom-branchingratio}. The model also includes the Gbb and Gtt decay topologies illustrated in Figure~\ref{fig:feynman-Gluino-topbottom-offshell}.  A limited set of 10 mass points were generated for this variable branching ratio model with $m_ {\gluino}$
varying from 1.5 TeV to 2.3 TeV and $m_ {\ninoone}$ varying from 1 GeV to 1 TeV.
 
\begin{figure*}[h]
\centering
\subfigure[]{\includegraphics[width=0.35\textwidth]{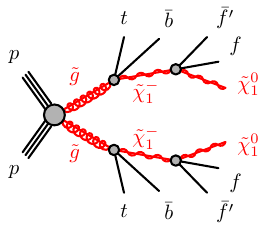}}
\subfigure[]{\includegraphics[width=0.35\textwidth]{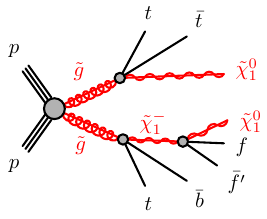}}
\subfigure[]{\includegraphics[width=0.35\textwidth]{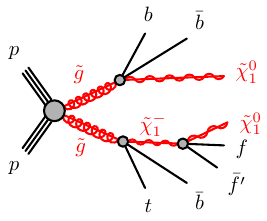}}
\subfigure[]{\includegraphics[width=0.35\textwidth]{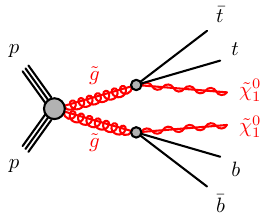}}
\caption{The additional decay topologies of the variable gluino branching ratio model in addition to the ones of Figure 1. (a) Both gluinos can decay as $\gluino \to t \bar{b} \chinoonem{}$ with $\chinoonem{} \to f\bar{f}' \ninoone$, or only one can with the other decaying as (b) $\gluino \to t\bar{t}\ninoone$ or (c) $\gluino \to b\bar{b}\ninoone$. (d) Finally, one gluino can decay as $\gluino \to t\bar{t}\ninoone$ and the other as $\gluino \to b\bar{b}\ninoone$.  The charge conjugate processes are implied.
The fermions originating from the $\chinoonepm$ decay are typically soft because the mass difference  between the $\chinoonepm$ and the $\ninoone$ is fixed to 2 GeV.
}\label{fig:feynman-Gluino-topbottom-branchingratio}
\end{figure*}

The technical implementation of the simulated samples produced from these models is described in Section~\ref{sec:samples}.
 
\section{ATLAS detector}
\label{sec:detector}
The ATLAS detector is a multipurpose particle physics detector with a forward-backward
symmetric cylindrical geometry and nearly 4$\pi$ coverage in solid angle.\footnote{ATLAS
uses a right-handed coordinate system with its origin at the nominal interaction point in
the centre of the detector. The positive $x$-axis is defined by the direction from the
interaction point to the centre of the LHC ring, with the positive $y$-axis pointing
upwards, while the beam direction defines the $z$-axis. Cylindrical coordinates $(r,\phi)$
are used in the transverse plane, $\phi$ being the azimuthal angle around the $z$-axis.
The pseudorapidity $\eta$ is defined in terms of the polar angle $\theta$ by $\eta=-\ln\tan(\theta/2)$.
Rapidity is defined as $y = 0.5 \ln  [(E + p_z )/(E - p_z ) ]$  where $E$ denotes the energy and $p_z$ is the component of the momentum along the beam direction.}
The inner tracking detector (ID) consists of silicon pixel and microstrip detectors
covering the pseudorapidity region $|\eta|<2.5$, surrounded by a transition radiation tracker,
which enhances electron identification in the region $|\eta|<2.0$.
Before the start of Run 2, the new innermost pixel layer, the insertable B-layer (IBL) \cite{ATLAS-TDR-19}, was inserted at a mean sensor radius of 3.3~cm.
The ID is surrounded by a thin superconducting solenoid providing an axial 2 T magnetic field and by
a fine-granularity lead/liquid-argon (LAr) electromagnetic calorimeter covering $|\eta|<3.2$.
A steel/scintillator-tile calorimeter provides coverage for hadronic showers in
the central pseudorapidity range ($|\eta|<1.7$).
The endcaps ($1.5<|\eta|<3.2$) of the hadronic calorimeter are made of LAr active layers with either copper or tungsten as the absorber material.
The forward region ($3.1 < |\eta| < 4.9$) is instrumented with a LAr calorimeter for both the EM and hadronic measurements.
A muon spectrometer with an air-core toroidal magnet system surrounds the calorimeters.
Three layers of high-precision tracking chambers
provide coverage in the range $|\eta|<2.7$, while dedicated fast chambers allow triggering in the region $|\eta|<2.4$.
The ATLAS trigger system \cite{TRIG-2016-01} consists of a hardware-based level-1 trigger followed by a software-based high-level trigger (HLT).
 
\section{Data and simulated event samples}
\label{sec:samples}
The data used in this analysis were collected by the ATLAS detector from $pp$ collisions produced by the
LHC at a centre-of-mass-energy of 13 \tev\ and 25 ns proton bunch spacing over the 2015 and 2016
data-taking periods. The full dataset corresponds to an integrated
luminosity of 36.1 \ifb\ after the application of beam, detector and data-quality requirements.
The uncertainty in the combined 2015+2016 integrated luminosity is 2.1\%.
It is derived, following a methodology similar to that detailed in
Ref.~\cite{DAPR-2013-01}, from a preliminary calibration of the
luminosity scale using $x$--$y$ beam-separation scans performed in August
2015 and May 2016. Events are required to pass an $\met$
trigger with thresholds of 70~\gev , 100~\gev\ and 110~\gev\ at the HLT level for
the 2015, early 2016 and late 2016 datasets, respectively. These triggers are fully efficient for events passing the
preselection defined in Section~\ref{sec:presel}, which requires the offline reconstructed $\met$ to exceed 200 GeV.
There are on average 24 inelastic $pp$ collisions (the interactions other than the hard scatter are referred to as ``pile-up'') in the dataset.
 
Samples of Monte Carlo (MC) simulated events are used to model the signal and background processes in this analysis, except
multijet processes, which are estimated from data. SUSY signal
samples in which each gluino decays into $b\bar{b}\ninoone$, $t\bar{t}\ninoone$, or  $t \bar{b} \chinoonem$ were generated with up to two
additional partons using \MGMCatNLO~\cite{Alwall:2014hca} v2.2.2 at leading order (LO) with the
NNPDF 2.3~\cite{Ball:2012cx} parton distribution function (PDF) set. These
samples were interfaced to \PYTHIA v8.186~\cite{Sjostrand:2007gs}
for the modelling of the parton showering, hadronisation and underlying event.
 
The dominant background in the signal regions is the production of
\ttbar\ pairs with additional high transverse momentum (\pt)
jets. For the generation of \ttbar\ and single top quarks in the $Wt$- and $s$-channels the
\POWHEGBOX~\cite{Alioli:2010xd}~v2 event generator with the CT10~\cite{Lai:2010vv} PDF set in the matrix
element calculations was used. Electroweak $t$-channel single-top-quark events were generated using the
\POWHEGBOX~v1 event generator. This event generator uses the
four-flavour scheme for the next-to-leading order (NLO) matrix elements
calculations together with the fixed four-flavour PDF set CT10f4. For
all processes involving top quarks, top-quark spin
correlations are preserved. In the $t$-channel, top quarks were decayed using MadSpin~\cite{Artoisenet:2012st}.
The parton shower, fragmentation, and the underlying event were
simulated using \PYTHIA v6.428~\cite{Sjostrand:2006za}
with the CTEQ6L1 PDF set~\cite{Pumplin:2002vw}. The $h_\mathrm{damp}$ parameter in \POWHEG,
which controls the \pt of the first additional emission beyond the Born level and thus regulates the \pt of the
recoil emission against the \ttbar\ system, was set to the mass of the top quark ($m_\mathrm{top} = $ 172.5 \gev).
All events with at least one leptonically decaying $W$ boson are
included. Single-top and \ttbar\ events in which all top quarks decay
hadronically do not contain sufficient \met\ to contribute significantly to the background.
 
Smaller backgrounds in the signal region come from the production of \ttbar\ pairs in association with $W/Z/h$
bosons and possibly additional jets, and production of $t\bar{t}t\bar{t}$,
$W/Z+$jets and $WW$/$WZ$/$ZZ$ (diboson) events.
Other potential sources of background, such as the
production of three top quarks or three gauge bosons, are expected to be negligible.
The production of \ttbar\ pairs in association with electroweak vector bosons $W$ and $Z$ was modelled by samples generated at LO using \MGMCatNLO~v2.2.2 and showered with \PYTHIA v8.186,
while samples to model ${\ttbar}H$ production were generated using
\MGMCatNLO~v2.2.1 and showered with \MYHERWIG++~\cite{Bahr:2008pv} 
v2.7.1. These samples are described in detail in Ref.~\cite{ATL-PHYS-PUB-2016-005}. \MGMCatNLO was also used to simulate the $\ttbar \ttbar$ production
and the showering was performed with \PYTHIA v8.186.
The $W/Z$+jets processes were simulated using the \SHERPA v2.2.0~\cite{Gleisberg:2008ta} event
generator, while \SHERPA v2.1.1 was used to simulate diboson
production processes. Matrix elements for the $W/Z$+jets and diboson processes were calculated using Comix~\cite{Gleisberg:2008fv}
and OpenLoops~\cite{Cascioli:2011va} and merged with the \SHERPA parton shower~\cite{Schumann:2007mg}
using the ME$+$PS$@$NLO prescription~\cite{Hoeche:2012yf}. The \SHERPA diboson sample cross-section was scaled
down to account for its use of $\alpha_{\mathrm{QED}} =1/129$ rather
than $1/132$, corresponding to the use of current Particle Data Group \cite{Patrignani:2016xqp} parameters,
as input to the $G_{\mu}$ scheme~\cite{ATL-PHYS-PUB-2016-002}. Samples generated using  \MGMCatNLO~v2.2.2 were produced
with the NNPDF 2.3 PDF set and $W/Z$+jets samples were generated with
the NNPDF 3.0 PDF set \cite{Ball:2014uwa}, while all other samples used CT10 PDFs.
 
For all samples, except the ones generated using \SHERPA, the \EVTGEN v1.2.0 program~\cite{EvtGen} was used to simulate
the properties of the bottom- and charm-hadron decays.
All \PYTHIA v6.428 samples used the  PERUGIA2012~\cite{Skands:2010ak} set of tuned parameters (tune)
for the underlying event, while \PYTHIA v8.186 and \MYHERWIG++ showering were run with the A14~\cite{ATL-PHYS-PUB-2014-021}
and UEEE5~\cite{ueee5} underlying-event tunes, respectively. In-time and out-of-time pile-up interactions
from the same or nearby bunch-crossings were simulated by overlaying additional $pp$ collisions generated by
\PYTHIA v8.186 using the A2 tune \cite{ATL-PHYS-PUB-2012-003} and the
MSTW2008LO parton distribution function set \cite{Martin:2009iq} on
top of the hard-scattering events.
Details of the sample generation and normalisation are
summarised in Table~\ref{t-MCsamples}. Additional samples with
different event generators and settings are used
to estimate systematic uncertainties in the backgrounds, as described in Section~\ref{sec:syst}.
 
All simulated event samples were passed through the full ATLAS detector simulation using \MYGEANT4~\cite{Agostinelli:2002hh},
with the exception of signal samples in which at least one gluino
decays as  $\gluino \to b\bar{b}\ninoone$ or $\gluino \to t \bar{b} \chinoonem$,
which were passed through a fast simulation that uses a
parameterisation for the calorimeter response \cite{ATL-PHYS-PUB-2010-013} and \MYGEANT4 for the
ID and the muon spectrometer.
The simulated events are reconstructed with the same algorithm as that used for data.
 
The signal samples are normalised using the best cross-section calculations at NLO in the strong coupling
constant, adding the resummation of soft gluon emission at next-to-leading-logarithm (NLL) accuracy
~\cite{Beenakker:1996ch,Kulesza:2008jb,Kulesza:2009kq,Beenakker:2009ha,Beenakker:2011fu}. The nominal cross-section
and the uncertainty are taken from an envelope of cross-section predictions using different PDF sets and factorisation
and renormalisation scales, as described in Ref.~\cite{Borschensky:2014cia}. The cross-section of gluino pair-production
in these simplified models is $14\pm 3$ fb for a gluino
mass of 1.5~\tev, falling to $1.0 \pm 0.3$ fb for 2~\tev\
mass gluinos. All background processes are normalised using the best available theoretical calculation for their
respective cross-sections. The order of this calculation in perturbative QCD (pQCD) for each process is listed in
Table~\ref{t-MCsamples}.  For \ttbar, the largest background, this corresponds to a
cross-section  of 831.8 pb.
 
Finally, contributions from multijet background are estimated from data using a procedure described in Ref.~\cite{Aad:2012fqa},
which performs a smearing of the jet response in data events with well-measured \met\ (so-called ``seed events''). The
response function is derived in Monte Carlo dijet events and is different for $b$-tagged and non-$b$-tagged jets.
 
\begin{table}[t]
 
\centering
{\footnotesize
\caption{List of event generators used for the different
processes. Information is given about the underlying-event
tunes, the PDF sets and the pQCD highest-order accuracy
used for the normalisation of the different samples.}
\label{t-MCsamples}
\begin{tabular}{ccccc}
\toprule
Process &    Event Generator                   & Tune & PDF set  &  Cross-section  \\
& + fragmentation/hadronisation  &      &          &  order \\
\midrule
{\textbf{SUSY signal}} &  \MGMCatNLO v2.2.2
& {\textsc A14} & {\textsc NNPDF2.3} & NLO+NLL~\cite{Beenakker:1996ch,Kulesza:2008jb,Kulesza:2009kq,Beenakker:2009ha,Beenakker:2011fu,Borschensky:2014cia} \\
& + \PYTHIA v8.186 & & & \\
\\
{\textbf{ \ttbar}} & \POWHEGBOX v2
& {\textsc PERUGIA2012} & {\textsc CT10} & NNLO+NNLL~\cite{Czakon:2011xx} \\
& + \PYTHIA v6.428 & & & \\
\\
{\textbf{ Single top}} & \POWHEGBOX v1 or v2
& {\textsc PERUGIA2012} & {\textsc CT10} & NNLO+NNLL~\cite{Kidonakis:2010tc,Kidonakis:2010ux,Kidonakis:2011wy}  \\
& + \PYTHIA v6.428 & & & \\
\\
{\textbf{ ${\ttbar}W$/${\ttbar}Z$/4-tops}} & \MGMCatNLO v2.2.2
& {\textsc A14} & {\textsc NNPDF2.3} & NLO~\cite{deFlorian:2016spz} \\
& + \PYTHIA v8.186 & & & \\
\\
{\textbf{ ${\ttbar}H$}} & \MGMCatNLO v2.2.1
& {\textsc UEEE5} & {\textsc CT10} & NLO~\cite{Heinemeyer:2013tqa} \\
& + \MYHERWIG++ v2.7.1 & & & \\
 
\\
{\textbf{ Diboson}}  & \SHERPA v2.1.1 & Default & {\textsc CT10} & NLO~\cite{ATL-PHYS-PUB-2016-002} \\
$WW$, $WZ$, $ZZ$ & & & & \\
\\
{\textbf{ $W$/${Z}+$jets}} & \SHERPA v2.2.0 & Default & {\textsc NNPDF3.0} & NNLO~\cite{Catani:2009sm}  \\
\bottomrule
\end{tabular}
}
\end{table}
 
 
\section{Event reconstruction}
\label{sec:object}
Interaction vertices from the proton--proton collisions are reconstructed from at least two tracks with
$\pt > 0.4~\gev$, and are required to be consistent with the beamspot envelope. The primary vertex is
identified as the one with the largest sum of squares of the transverse momenta from associated
tracks ($\sum{|p_{\mathrm{T,track}}|^{2}}$)~\cite{ATL-PHYS-PUB-2015-026}.
 
Basic selection criteria are applied to define candidates for electrons, muons and jets in the event. An
overlap removal procedure is applied to these candidates to prevent double-counting.
Further requirements are then made to select the final signal leptons and jets
from the remaining candidates. The details of the candidate selections and of the overlap removal procedure
are given below.
 
Candidate jets are reconstructed from three-dimensional topological energy clusters~\cite{PERF-2014-07} in the calorimeter using the anti-$k_{t}$ jet algorithm~\cite{Cacciari:2008gp, Cacciari:2011ma} with a radius parameter of 0.4 (small-$R$ jets).
Each topological cluster is calibrated to the electromagnetic scale response prior to jet reconstruction.
The
reconstructed jets are then calibrated to the particle level by the application of a jet energy scale
(JES) derived from $\sqrt{s}= 13~\tev$ data and simulations~\cite{Aaboud:2017jcu}.
Quality criteria are imposed to reject events that contain at least one jet arising from non-collision sources
or detector noise~\cite{ATLAS-CONF-2015-029}. Further selections are applied
to reject jets that originate from pile-up interactions by means of a multivariate algorithm using information
about the tracks matched to each jet~\cite{PERF-2014-03}. Candidate jets are
required to have \pt$>$~20 \gev\ and $|\eta| < 2.8$. After resolving overlaps with electrons
and muons, selected jets are required to satisfy the stricter requirement of \pt$>$~30~\gev.
 
A jet is tagged as a $b$-jet candidate by means of a multivariate algorithm using information about the impact
parameters of inner detector tracks matched to the jet, the presence of displaced secondary vertices, and the
reconstructed flight paths of $b$- and $c$-hadrons inside the jet~\cite{PERF-2012-04,ATL-PHYS-PUB-2016-012}.
The $b$-tagging working point corresponding to an efficiency of 77\% to identify $b$-jets with \pt$>$~20 \gev, as
determined from a sample of simulated \ttbar events, is found to be optimal for the statistical significance of this search.
The corresponding rejection factors against jets originating from $c$-quarks, $\tau$-leptons and light
quarks and gluons in the same sample at this working point are 6, 22 and 134, respectively.
 
After resolving the overlap with leptons, the candidate small-$R$ jets are re-clustered~\cite{Nachman:2014kla} into
large-$R$ jets using the anti-$k_{t}$ algorithm with a radius parameter of 0.8.
The calibration from the input small-$R$ jets propagates directly to the re-clustered jets.
These re-clustered jets are then trimmed~\cite{Krohn:2009th,PERF-2012-02,Nachman:2014kla, ATLAS-CONF-2017-064} by removing subjets whose
\pt falls below 10\% of the \pt\ of the original re-clustered jet.
The resulting large-$R$ jets are required to have \pt$>$~100 \gev\ and $|\eta|<$~2.0. When it is not explicitly
stated otherwise, the term ``jets'' in this paper refers to small-$R$ jets.
 
Electron candidates are reconstructed from energy clusters in the electromagnetic calorimeter and inner
detector tracks and are required to satisfy a set of ``loose'' quality criteria~\cite{PERF-2013-05,ATLAS-CONF-2016-024}.
They are also required to have $|\eta|<$~2.47. Muon candidates are reconstructed from matching tracks in
the inner detector and muon spectrometer. They are required to meet ``medium'' quality criteria,
as described in Ref.~\cite{PERF-2015-10}, and to have $|\eta|<$~2.5. All electron and muon
candidates must have \pt$>$~20 \gev.
 
Leptons are selected from the candidates that survive the overlap removal procedure if they fulfil a requirement on the scalar
sum of \pt of additional inner detector tracks in a cone around the lepton track. This isolation requirement is defined to ensure
a flat efficiency of around 99\% across the whole electron transverse energy and muon transverse momentum ranges. The angular
separation between the lepton and the $b$-jet ensuing from a semileptonic top quark decay narrows as the \pt of the top quark
increases. This increased collimation is accounted for by setting the radius of the isolation cone to
min(0.2, 10~\gev/$p_{\mathrm{T}}^{\mathrm{lep}}$), where $p_{\mathrm{T}}^{\mathrm{lep}}$ is the lepton \pt expressed in~\gev.
Selected electrons are further required to meet the ``tight'' quality criteria~\cite{PERF-2013-05,ATLAS-CONF-2016-024}.
Electrons (muons) are matched to the primary vertex by requiring the transverse impact parameter $d_{0}$ of the associated ID track to
satisfy $|d_{0}|/\sigma_{d_{0}}<$~5 (3), where $\sigma_{d_{0}}$ is the measured uncertainty of $d_{0}$,
and the longitudinal impact parameter $z_{0}$ to satisfy $|z_{0}\sin\theta|<$~0.5 mm.\footnote{Both the transverse and longitudinal impact
parameters are defined with respect to the selected primary vertex.} In addition, events
containing one or more muon candidates with $|d_{0}|$ ($|z_{0}|$) $>$ 0.2~mm (1~mm) are rejected to suppress
cosmic rays.
 
Overlaps between candidate objects are removed sequentially. Firstly, electron candidates that
lie a distance\footnote{${\Delta}R = \sqrt{(\Delta y)^{2} + (\Delta\phi)^{2}}$  defines the distance in rapidity $y$ and
azimuthal angle $\phi$.} ${\Delta}R < 0.01$ from muon candidates are removed to suppress contributions from muon bremsstrahlung.
Overlaps between electron and jet candidates are resolved next, and finally, overlaps between remaining
jets and muon candidates are removed.
 
Overlap removal between electron and jet candidates aims to resolve two sources of ambiguity: it is designed, firstly,
to remove jets that are formed primarily from the showering of a prompt electron and, secondly, to remove electrons
that are produced in the decay chains of hadrons.
Consequently, any non-$b$-tagged jet whose axis lies ${\Delta}R < 0.2$ from an electron is discarded. Electrons
with $\et < 50$~\gev\ are discarded if they lie ${\Delta}R < 0.4$ from the axis of any remaining jet and the corresponding jet is
kept. For higher-\et\ electrons, the latter removal is performed using a threshold of
${\Delta}R = \mathrm{min}(0.4, 0.04 + 10~\gev/\et)$ to increase the acceptance for events with collimated top quark decays.
 
The procedure to remove overlaps between muon and jet candidates is designed to remove those muons
that are likely to have originated from the decay of hadrons and to retain the overlapping jet.
Jets and muons may also appear in close proximity when the jet results from high-\pt muon bremsstrahlung, and
in such cases the jet should be removed and the muon retained. Such jets are characterised by having very
few matching inner detector tracks. Therefore, if the angular distance ${\Delta}R$ between a muon and a jet is
lower than 0.2, the jet is removed if it is not $b$-tagged and has fewer than three matching
inner detector tracks.
Like the electrons, muons with \pt below (above) 50~\gev\ are subsequently discarded if they lie
within $\Delta R = 0.4$ ($\Delta R = \min(0.4, 0.04 + 10~\gev/\pt)$) of any remaining jet.

The missing transverse momentum ($\met$) in the event is defined as the magnitude of the negative
vector sum ($\vec{\pt}^{\mathrm{miss}}$) of the transverse momenta of
all selected and calibrated objects in the event, with an extra term added to account for energy deposits that
are not associated with any of these selected objects. This ``soft'' term is calculated from
inner detector tracks matched to the primary vertex to make it more resilient to contamination from
pile-up interactions ~\cite{ATL-PHYS-PUB-2015-027,ATL-PHYS-PUB-2015-023} .

Corrections derived from data control samples are applied to simulated events to account for differences between
data and simulation in the reconstruction efficiencies, momentum scale and resolution of leptons, in the efficiency
and fake rate for identifying $b$-jets, and in the efficiency for rejecting jets originating from pile-up
interactions.

 
\section{Event selection}
\label{sec:presel}
The event selection criteria are defined based on kinematic requirements for the objects defined
in Section~\ref{sec:object}. Other discriminating event-based
variables, described in Section~\ref{sec:variables}, are used to further reject the
background. Two sets of preselection criteria targeting
the 0-lepton and the 1-lepton channels are presented in
Section~\ref{sec:preselection}. The modelling of the data in these
regions is also discussed in that section. The general analysis
strategy and the treatment of background sources is presented in
Section \ref{sec:strategytreatment}. Finally, the event selection for the cut-and-count and
multi-bin analyses are discussed in Sections \ref{sec:cutandcount} and
\ref{sec:multibin}, respectively.
 
\subsection{Discriminating variables}
\label{sec:variables}
 
The effective mass variable (\meff) is defined as:
\begin{equation*}
\meff = \sum_{i} \pt^{\mathrm{jet}_{i}}  + \sum_{j} \pt^{\ell_j}  + \met,
\end{equation*}
where the first and second sums are over the selected jets ($\njet$) and leptons ($N_\mathrm{lepton}$),
respectively. It typically has a much higher value in pair-produced gluino
events than in background events.
 
In regions with at least one selected lepton, the transverse mass
$\mt$ composed of the $\pt$ of the leading selected lepton ($\ell$)
and $\met$ is defined as:
\begin{equation*}
\mt = \sqrt{2\pt^\ell \met\{1-\cos[\Delta\phi(\vec{p}_{\mathrm{T}}^{\mathrm{miss}},\vec{p}_{\mathrm{T}}^{\ell})]\}}.
\end{equation*}
It is used to reduce the $\ttbar$  and $W$+jets background events
in which a $W$ boson decays leptonically.
Neglecting resolution effects, the $\mt$ distribution for these
backgrounds has an expected upper bound corresponding to the $W$ boson mass
and typically has higher values for Gtt events. Another useful transverse mass variable is $\mtb$, the minimum
transverse mass formed by $\met$ and any of the three highest-\pt $b$-tagged jets in the event:
\begin{equation*}
\mtb =  \mathrm{min}_{i\leq 3}  \left ( \sqrt{2\pt^{b\textrm{-jet}_i}
\met\{1-\cos[\Delta\phi(\vec{p}_{\mathrm{T}}^{\mathrm{miss}},\vec{p}_{\mathrm{T}}^{b\textrm{-jet}_i} )]\}} \right ).
\end{equation*}
The  $\mtb$ distribution
has an expected upper bound corresponding to the top quark mass for $\ttbar$ events with a semileptonic
top quark decay, while peaking at higher values for Gbb
and Gtt events.
 
Another powerful variable is the total jet mass variable, defined as:
 
\begin{equation*}
\mjsum = \sum_{i\leq 4} m_{J,i},
\end{equation*}
 
where $m_{J,i}$ is the mass of the large-radius re-clustered jet $i$ in
the event. The decay products of a hadronically decaying boosted top
quark can be reconstructed in a single large-radius re-clustered
jet, resulting in a jet with a high mass. This variable typically has
larger values for Gtt events than for background events. This is because
Gtt events
contain as many as four hadronically decaying top quarks while the
background is dominated by \ttbar events with one or
two semileptonic top quark decays.
 
The requirement of a selected lepton, with the
additional requirements on jets, $\met$ and event variables described above,
makes the multijet background negligible for the $\ge$ 1-lepton signal regions. For the 0-lepton signal regions, the
minimum azimuthal angle $\dphimin$ between $\vec{p}_{\mathrm{T}}^{\mathrm{miss}}$ and the
$\pt$ of the four leading small-$R$ jets in the
event, defined as:
\begin{equation*}
\dphimin = \textrm{min}_{i\leq 4} \left(|\phi_{\mathrm{jet}_i} - \phi_{\vec{p}_{\mathrm{T}}^{\mathrm{miss}}}| \right),
\end{equation*}
is required to be greater than 0.4.
This requirement supresses the multijet background, which can
produce events with large $\met$ if containing
poorly measured jets or neutrinos emitted close to the axis of a jet.
A similar variable, denoted $\dphilead$, is also used in the Gbb signal regions targeting small mass differences
between the gluino and the neutralino, allowing the identification of
events containing a high-\pt\ jet coming from initial-state radiation (ISR) and recoiling against the gluino pair. It is defined
as the absolute value of the azimuthal angle separating the $\pt$ of the leading jet and $\vec{p}_{\mathrm{T}}^{\mathrm{miss}}$, and is expected to
have larger values for the targeted signal than for the background.
 
\subsection{Modelling of the data}
\label{sec:preselection}
 
Preselection criteria in the 0-lepton and
1-lepton channels
require $\met > 200$~\gev, in addition to the
$\met$ trigger requirement, and at least four jets of which at least two
must be $b$-tagged. The 0-lepton (1-lepton) channel requires the event to contain no (at least one) selected lepton.

In this analysis, correction factors need to be extracted to account for shape discrepancies in the \meff\ spectrum
between the data and the expected background for the 1-lepton preselection sample. These factors are
defined as the ratio of the number of observed events to the predicted number of background events in a given
\meff\ bin, in a signal-depleted region. This region is defined by applying the 1-lepton preselection criteria
and requiring exactly two $b$-tagged jets and $\mtb < 140$~\GeV.
This kinematic reweighting leads to correction factors ranging from 0.7 to 1.1. They are applied to the
background prediction and the full size of the correction is taken as an uncertainty for both the background and
signal events.
 
Figures~\ref{fig:presel_0l} and \ref{fig:presel_1l} show the multiplicity of selected jets and
$b$-tagged jets, the distributions of \met, \meff, and \mjsum\ for events passing the 0-lepton
or the 1-lepton preselection, respectively. Figure \ref{fig:presel_0l} (\ref{fig:presel_1l}) also
displays the distribution of \mtb\ (\mt) in the 0-lepton (1-lepton)
channel. The correction described above is applied in the 1-lepton
channel.
The uncertainty bands include the statistical and experimental systematic uncertainties, as described
in Section~\ref{sec:syst}, but not the theoretical uncertainties in the background modelling.
 
The data and the predicted background are found to agree reasonably well at the preselection level
after the kinematic reweighting described above.
A discrepancy between data and prediction is observed for the number of $b$-tagged jets, but it has a
negligible impact on the background estimate after the renormalisation of the simulation in dedicated control
regions with the same $b$-tagged jets requirements as the signal regions, as described in
Sections~\ref{sec:cutandcount} and \ref{sec:multibin}. Example signal models with enhanced cross-sections are overlaid for comparison.
 
\begin{figure}[htbp]
\centering
\includegraphics[width=0.490\textwidth]{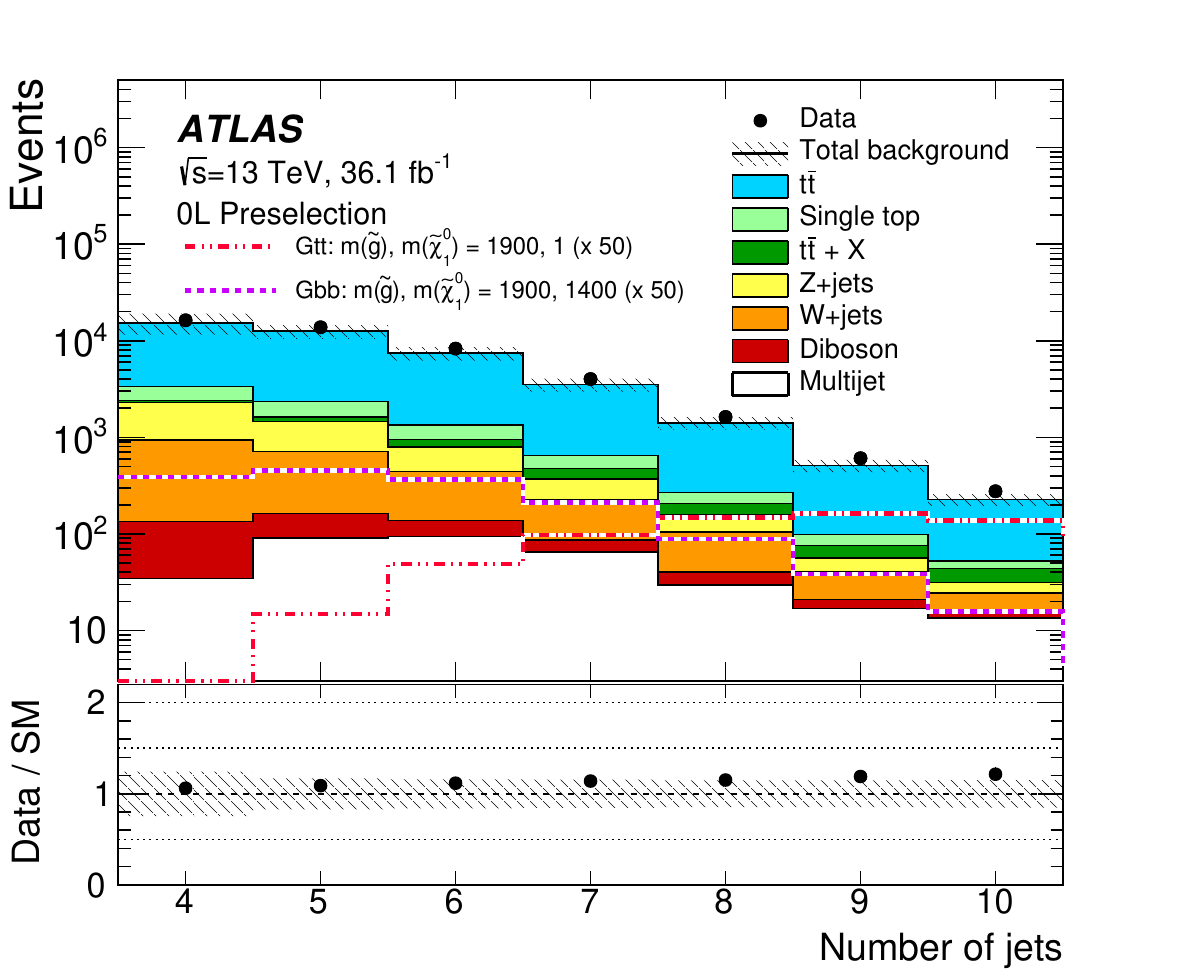}
\includegraphics[width=0.490\textwidth]{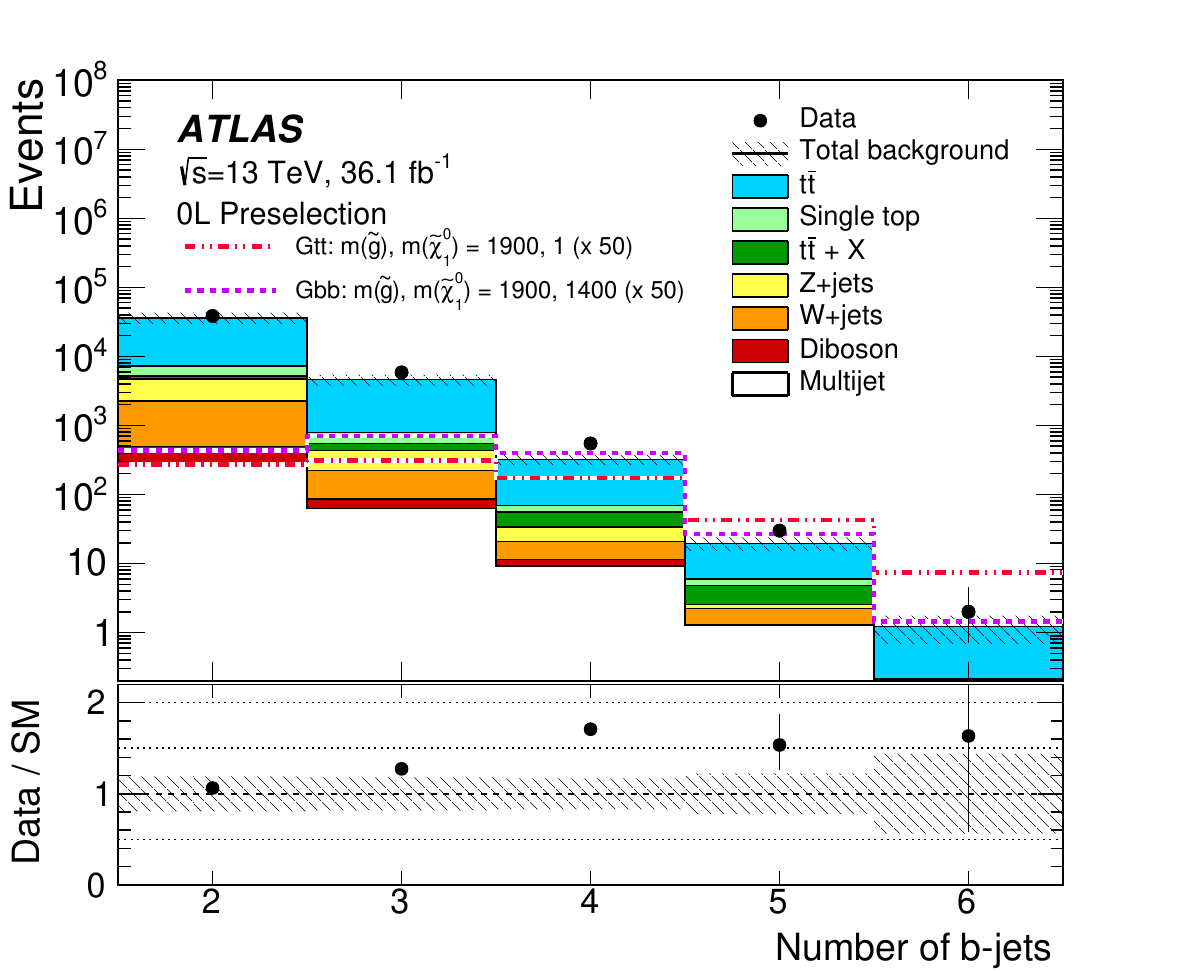}
\includegraphics[width=0.490\textwidth]{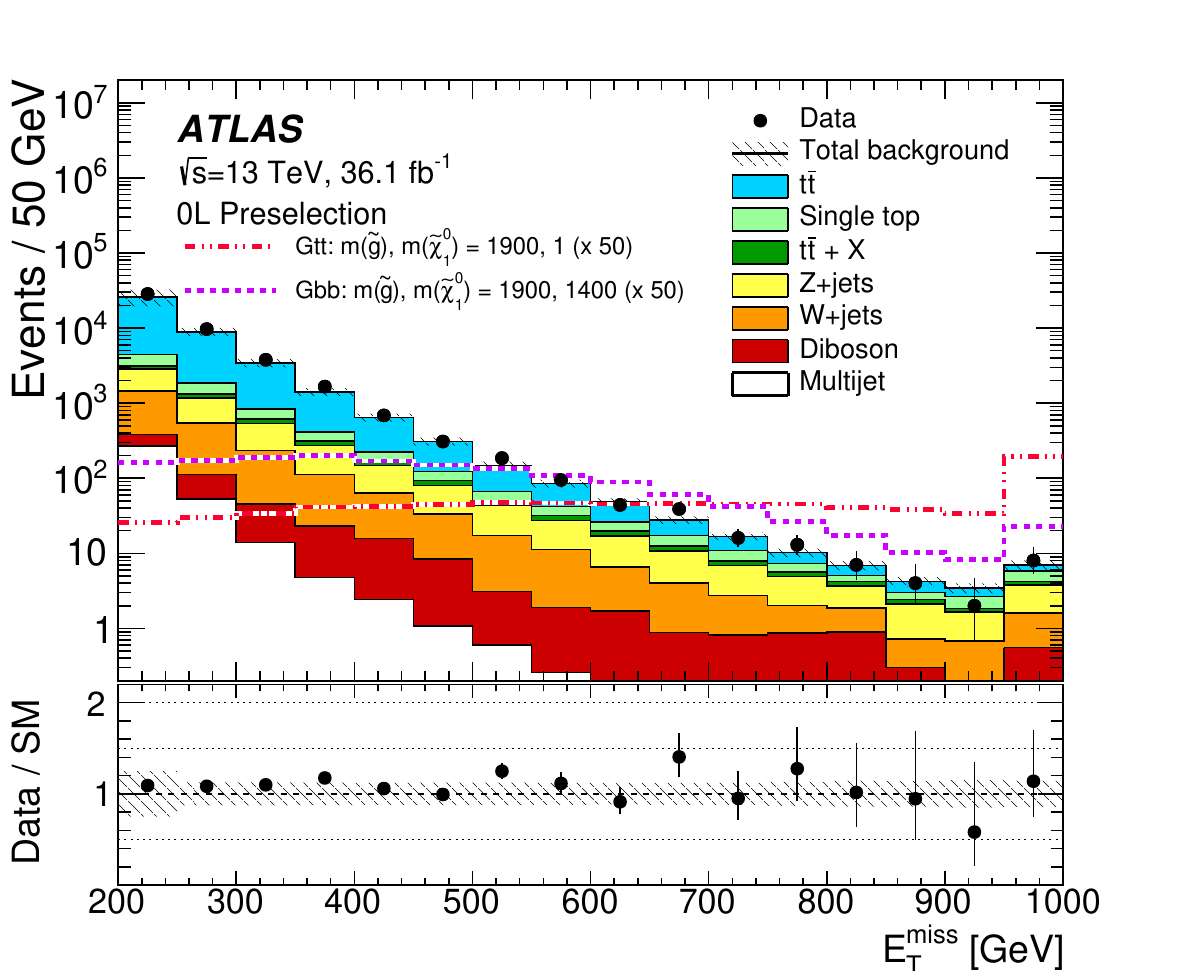}
\includegraphics[width=0.490\textwidth]{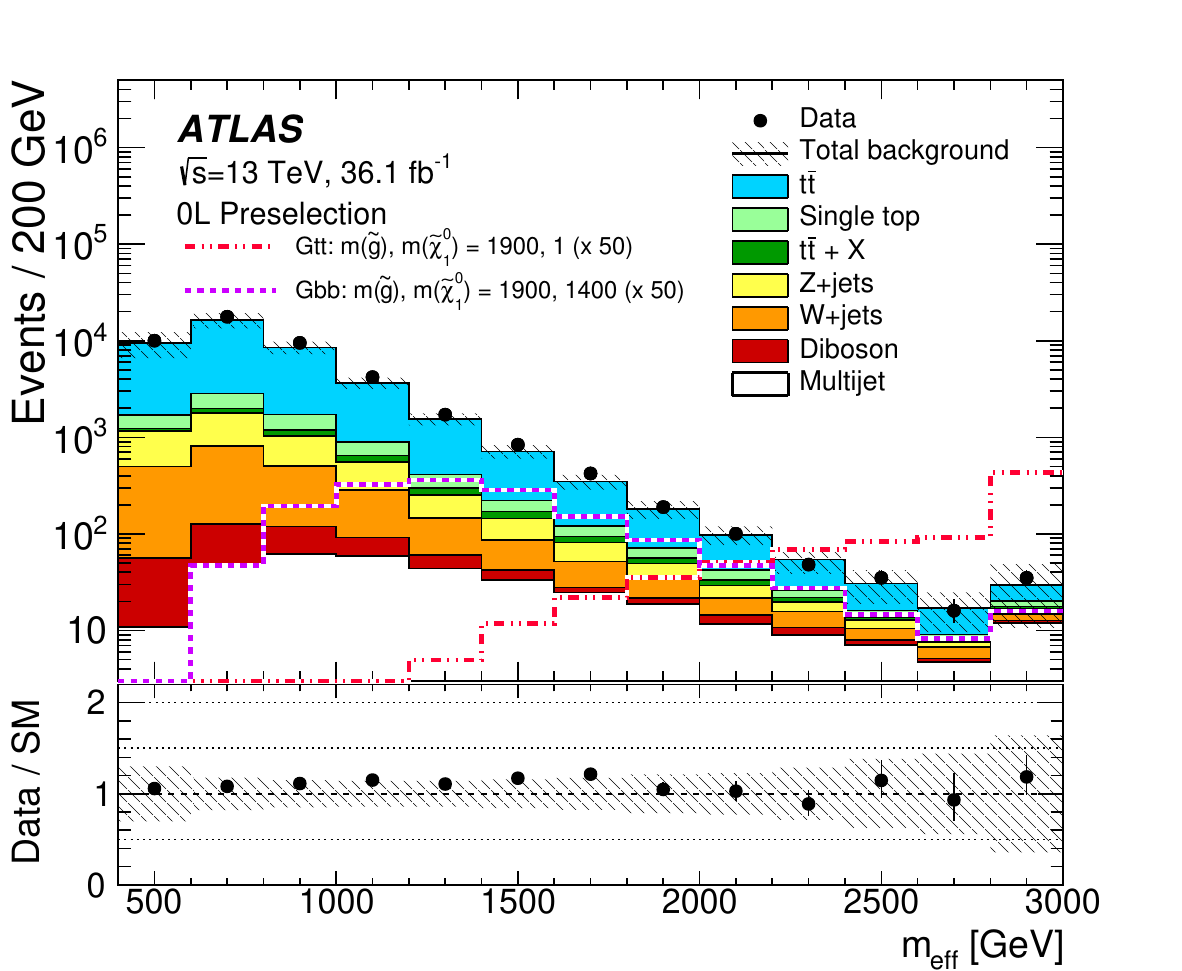}
\includegraphics[width=0.490\textwidth]{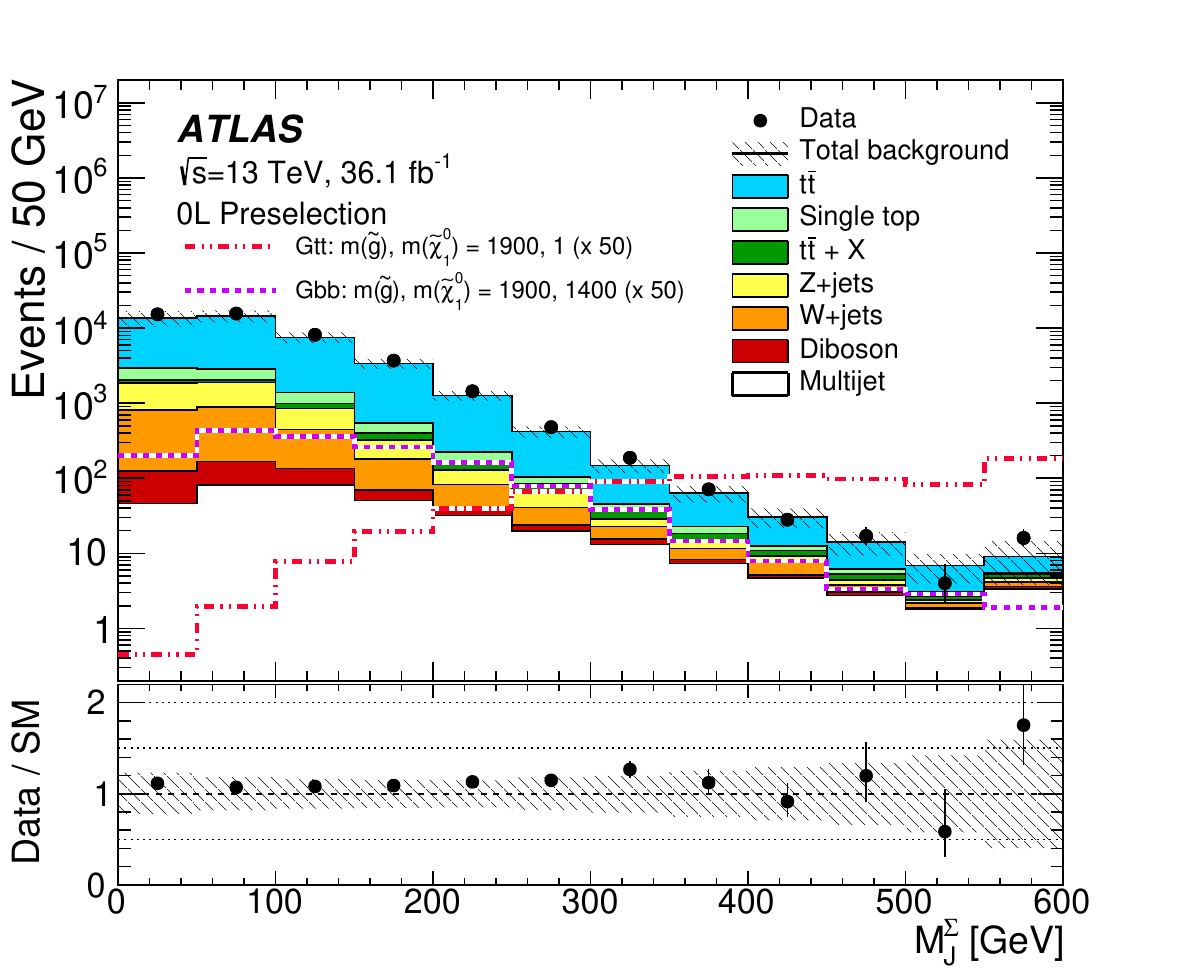}
\includegraphics[width=0.490\textwidth]{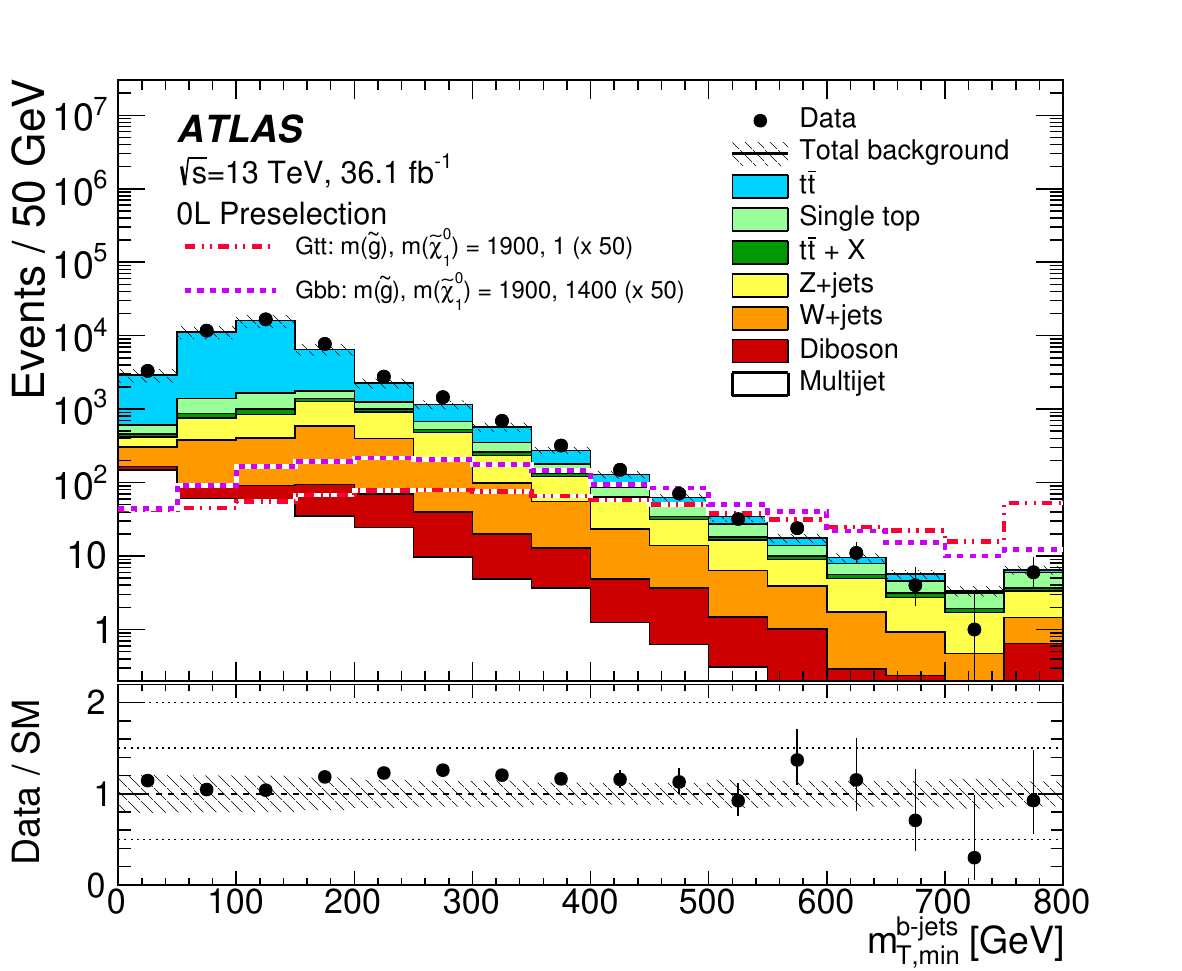}
\caption{Distributions of (top-left) the number of selected
jets ($\njet$), (top-right) the number of selected
$b$-tagged jets, (centre-left) \met, (centre-right) \meff,
(bottom-left) \mjsum\ and (bottom-right) \mtb\ for events passing the 0-lepton preselection criteria. The statistical and experimental
systematic uncertainties (as  defined in
Section~\ref{sec:syst}) are included in the
uncertainty band. The last bin includes overflow events.
The lower part of each figure shows
the ratio of data to the background prediction. All backgrounds (including  $\ttbar$) are
normalised using the best available theoretical calculation described in Section~\ref{sec:samples}.
The background category $\ttbar+X$ includes $\ttbar W/Z$, $\ttbar h$ and $\ttbar \ttbar$ events.
Example signal models with cross-sections enhanced by a factor of 50 are overlaid for comparison.
}
\label{fig:presel_0l}
\end{figure}
 
\begin{figure}[htbp]
\centering
\includegraphics[width=0.490\textwidth]{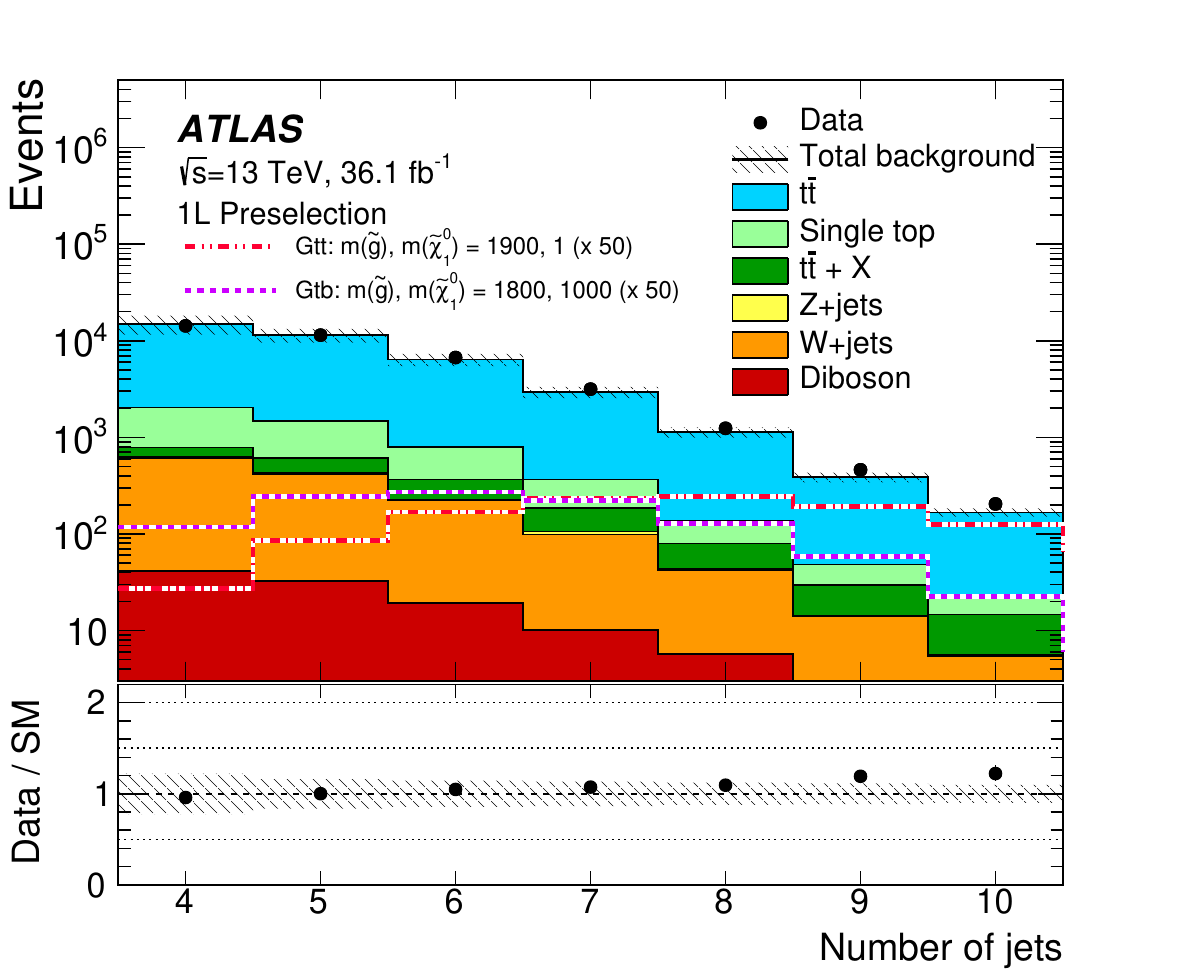}
\includegraphics[width=0.490\textwidth]{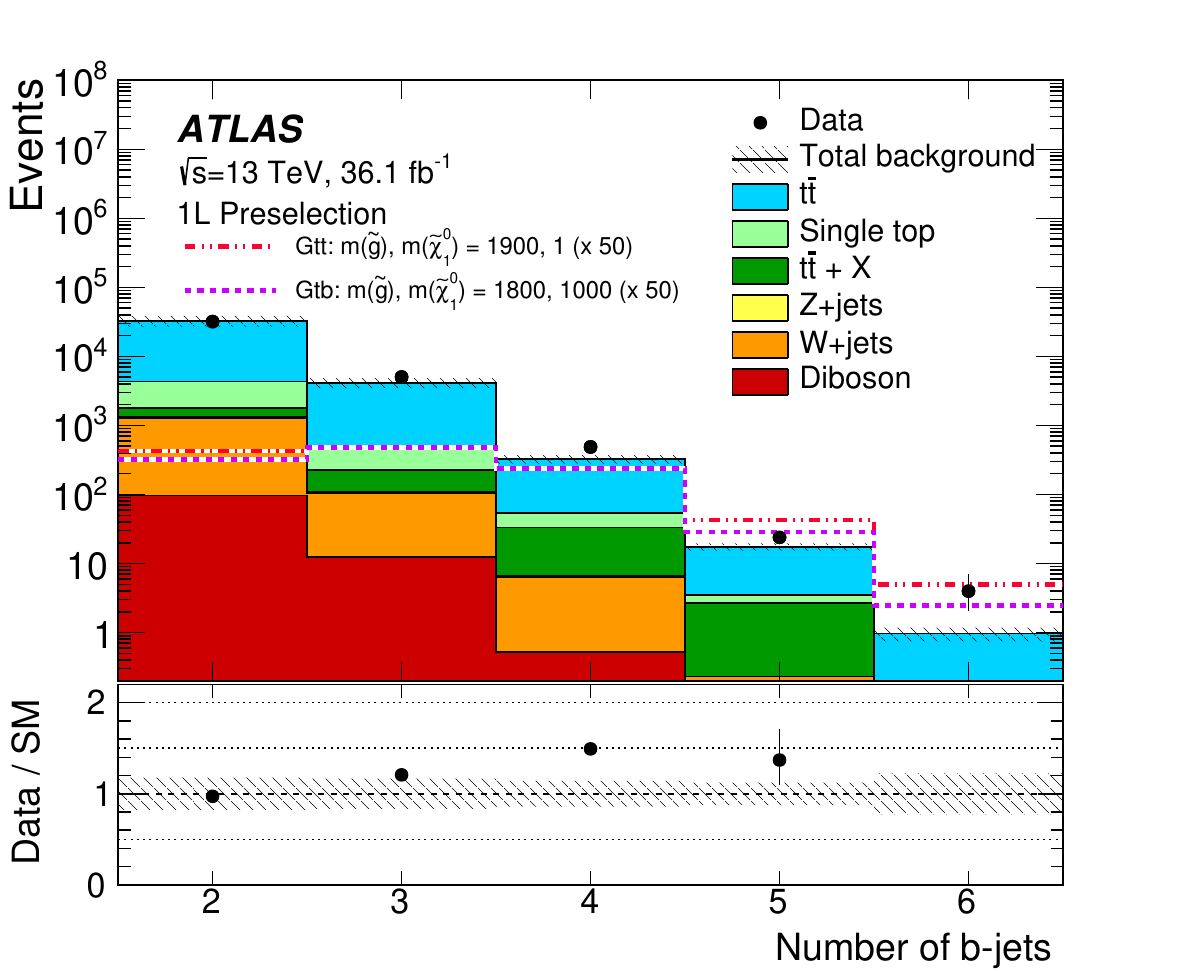}
\includegraphics[width=0.490\textwidth]{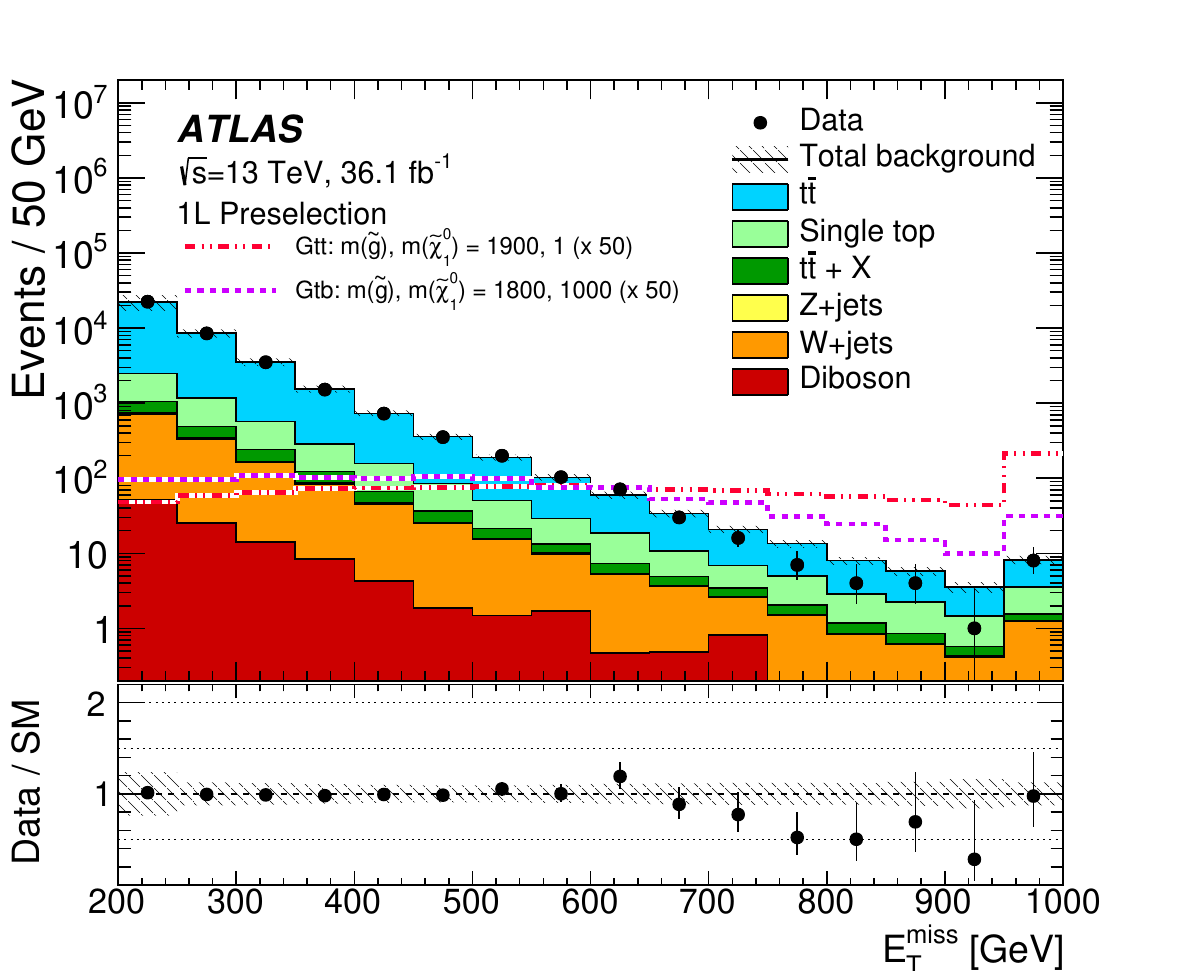}
\includegraphics[width=0.490\textwidth]{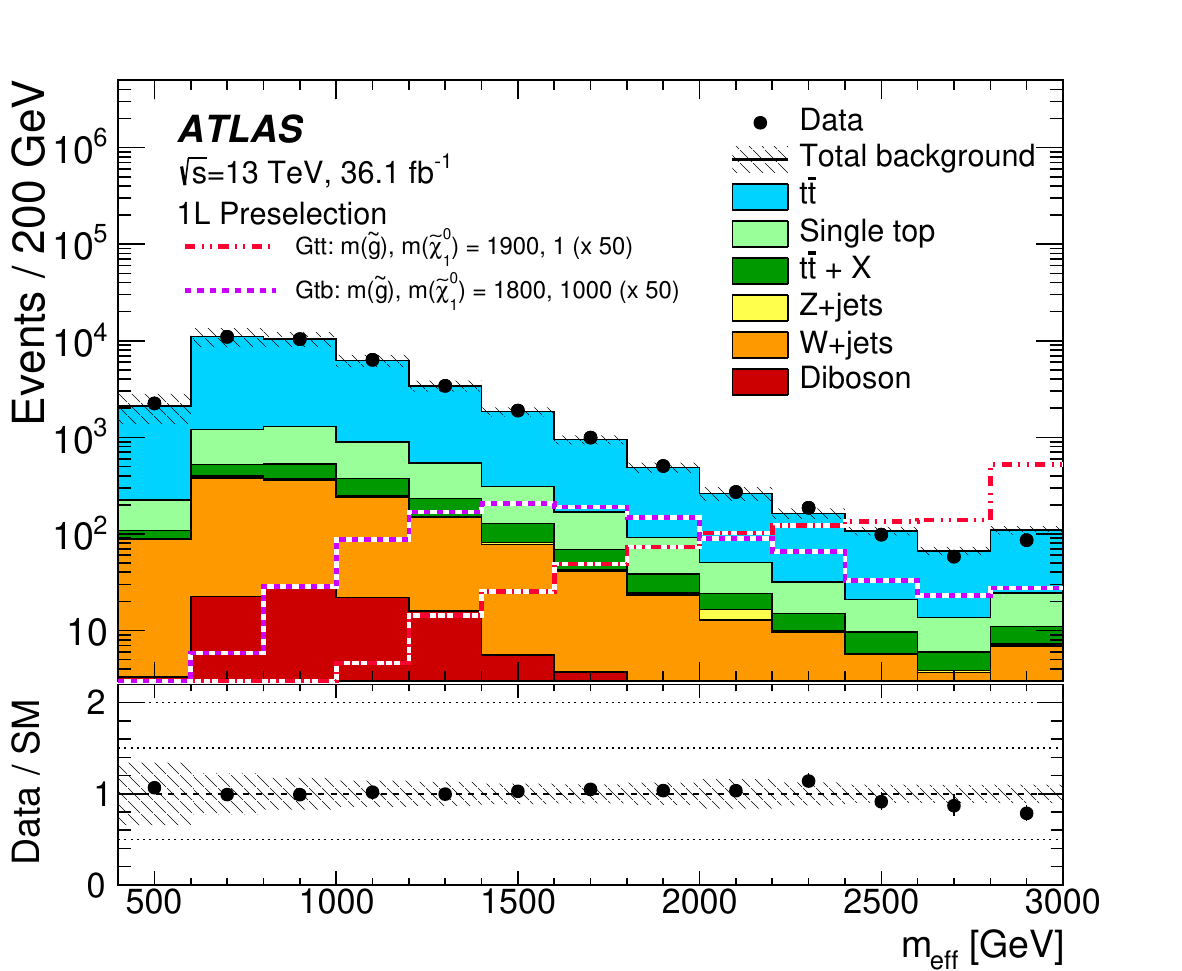}
\includegraphics[width=0.490\textwidth]{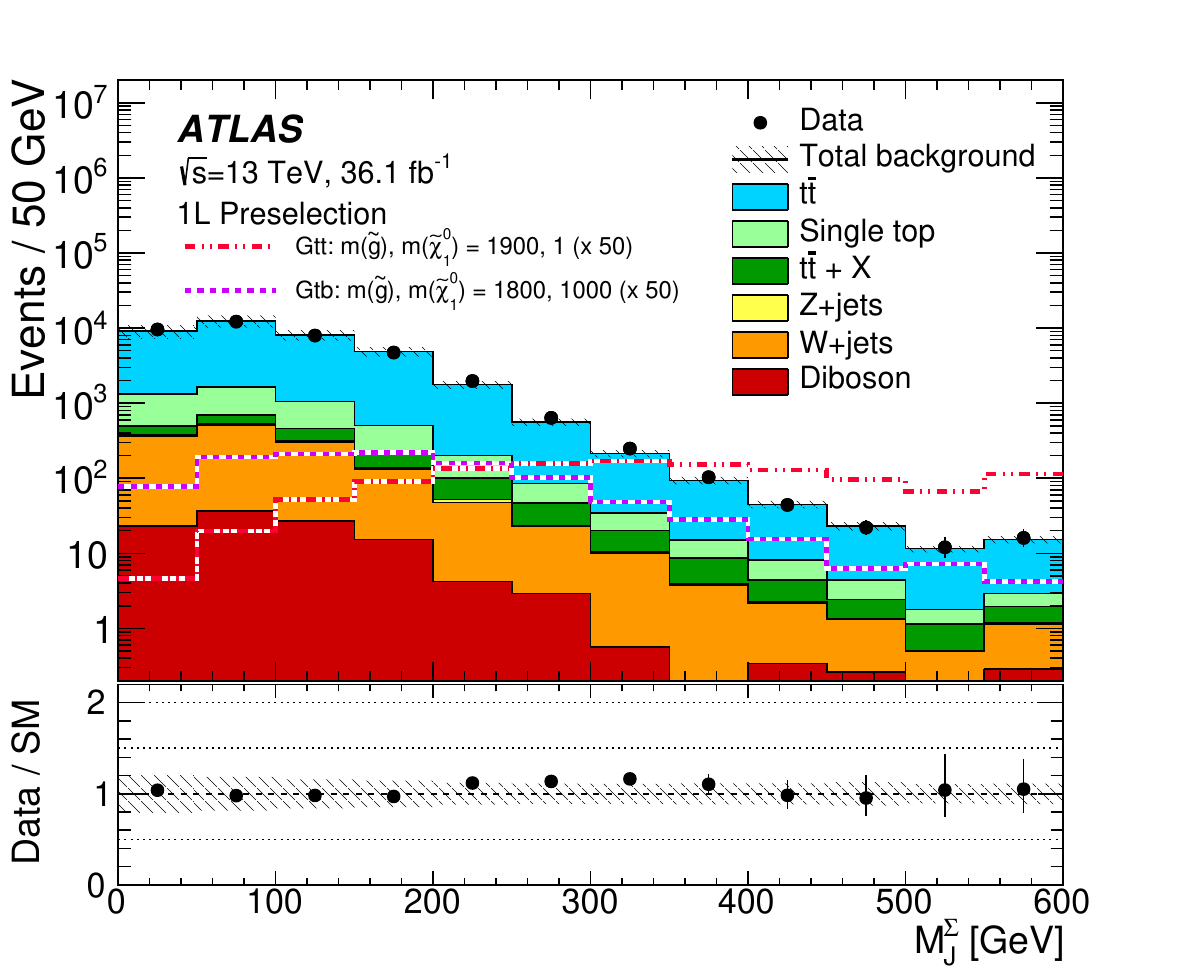}
\includegraphics[width=0.490\textwidth]{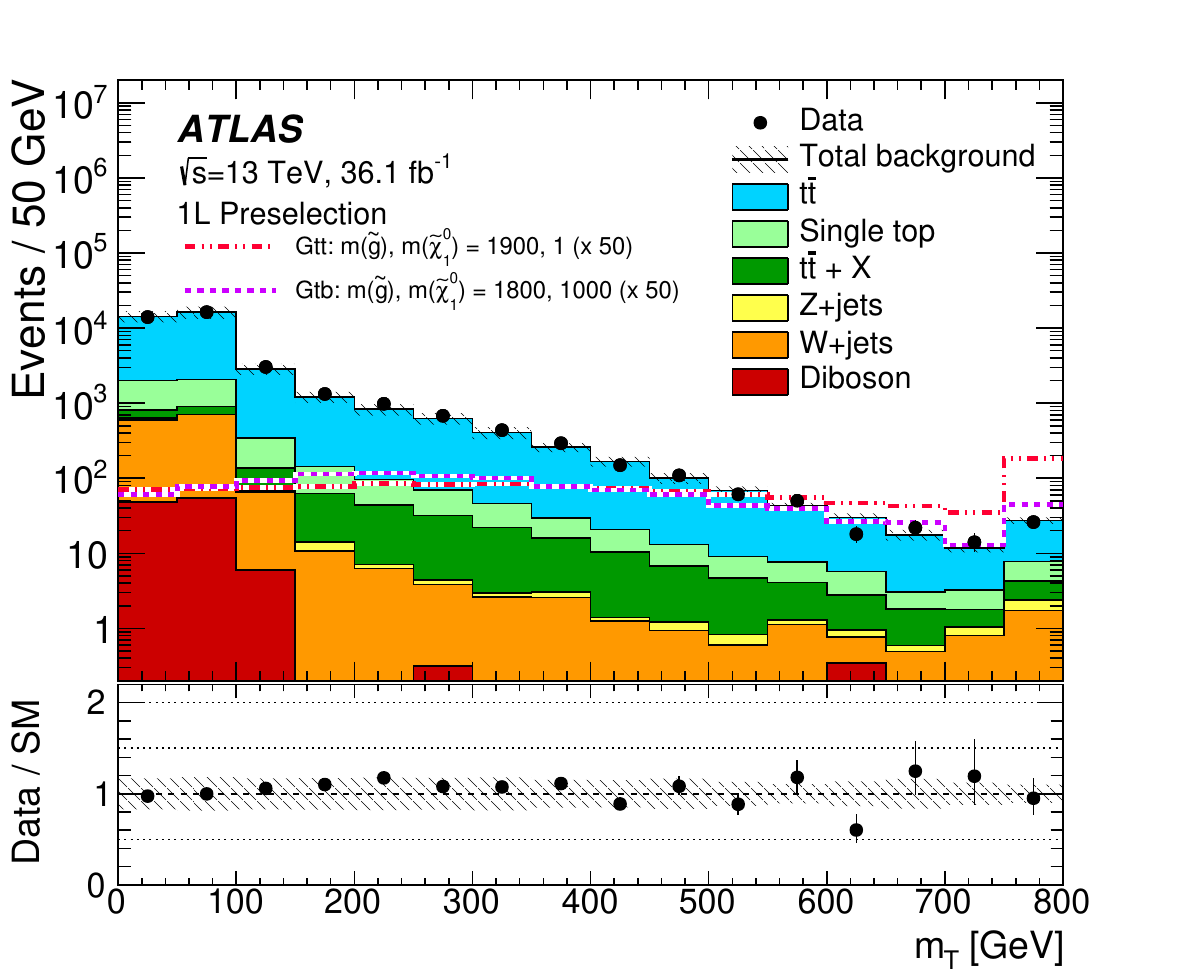}
\caption{Distributions of (top-left) the number of selected
jets ($\njet$), (top-right) the number of selected $b$-tagged jets, (centre-left) \met, (centre-right) \meff,
(bottom-left) \mjsum\ and (bottom-right) \mt\ for events passing the
1-lepton preselection criteria, after applying the
kinematic reweighting to the \meff\ distribution
described in the text.
The statistical and experimental
systematic uncertainties (as  defined in
Section~\ref{sec:syst}) are included in the
uncertainty band. The last bin includes overflow events.
The lower part of each figure shows
the ratio of data to the background prediction. All backgrounds (including  $\ttbar$) are
normalised using the best available theoretical calculation described in Section~\ref{sec:samples}.
The background category $\ttbar+X$ includes $\ttbar W/Z$, $\ttbar h$ and $\ttbar \ttbar$ events.
Example signal models with cross-sections enhanced by a factor of 50 are overlaid for comparison.
}
\label{fig:presel_1l}
\end{figure}
 
\subsection{Analysis strategy and background treatment}
\label{sec:strategytreatment}
 
In order to enhance the sensitivity to the various signal benchmarks described in Section \ref{sec:susy_sig},
multiple signal regions (SRs) are defined. The main background in all
these regions is the production of a \ttbar\ pair in association with heavy- and light-flavour
jets. A normalisation factor for this background is extracted  for each individual SR from a data control region (CR)
that has comparable background composition and kinematics.
This is ensured by keeping the kinematic
requirements similar in the two regions.
The CRs and SRs are defined to be mutually exclusive.
Signal contributions in the CRs are suppressed by inverting or relaxing some requirements on the
kinematic variables (e.g. \mt\ or \mtb), leading to a signal contamination in the CRs of 6\% at most.
The $\ttbar$ normalisation is cross-checked
in validation regions (VRs) that share similar background composition, i.e.\ jet and lepton flavours,
with the SR. The signal contamination in the VRs is found to be lower than 30\% for
benchmark signal mass points above the already excluded mass
range. The \ttbar\ purity is superior to 73\% and 53\% in the CRs and
VRs, respectively.
 
The non-\ttbar\ backgrounds mainly consist of single-top, $W$+jets,
$Z$+jets, \ttbar+$W/Z/h$, $t\bar{t}t\bar{t}$ and diboson events.
Their normalisation is taken from the
simulation normalised using the best available theory prediction.
The multijet background is found to be very small or negligible in all regions. It
is estimated
using a procedure described in Ref. \cite{Aad:2012fqa}, in which  the jet
response is determined from
simulated dijet events. This response function is then used to smear
the jet response in low-\met\ events.
The jet response is cross-checked with data where the \met\ can be unambiguously attributed to the
mismeasurement of one of the jets.
 
Two analysis strategies are followed, and different SR sets are defined for each:
\begin{itemize}
\item A \textbf{cut-and-count} analysis, using partially overlapping single-bin SRs,
optimised to maximise the expected discovery power for
benchmark signal models, and allowing for reinterpretation of
the results.
The SRs are defined to probe the existence of a signal or to assess
model-independent upper limits on the number of signal events.
\item A \textbf{multi-bin} analysis, using a set of non-overlapping SRs and CRs that
are combined to strengthen the exclusion limits on the targeted signal benchmarks.
This set of regions is used to assess model-dependent interpretations of the various signal models.
\end{itemize}

\subsection{Cut-and-count analysis}
\label{sec:cutandcount}

The SRs are named in the form SR-{\textit X}-{\textit
Y}L-{\textit Z}, where {\textit X} indicates the target model,
{\textit Y} indicates the number of leptons and {\textit Z} labels the
type of region targeted.
The  cut-and-count regions
labelled B (for ``boosted'') are optimised for signals with a large mass difference between the gluino and the neutralino
($\msplit \gtrsim 1.5$~\tev), possibly leading to highly boosted
objects in the final state. Conversely, regions C (for ``compressed'')
primarily
focus on signals for which the gluino decay products are softer due to the small \msplit ($\msplit \lesssim~300$~\gev).
Regions M (for ``moderate'') target intermediate values of
\msplit. SRs targeting the Gtt model in the 1- and 0-lepton channels
are presented in Table \ref{tab:GttEvsel}.
 
In the 1-lepton channel,
these regions differ mainly in their kinematic selections thresholds: \meff, \met\ and \mjsum\ selections are
relaxed when going from region B to C to improve the acceptance for softer signals. The resulting background increase is compensated for by tightening
the requirements on the number of ($b$-tagged) jets or \mtb.
CRs constraining the \ttbar\ background are defined in the low-\mt\ region to remove overlaps
with the SRs. The requirements on \mtb\ are removed, and the selections on kinematic variables
are relaxed to ensure at least about 10 events in each CR.
The requirement of an exclusive jet multiplicity permits the definition of VRs kinematically close to the SRs and mutually exclusive to both the CRs and SRs.
VR-\mt\ validates the background prediction in the high-\mt\ region. It is kept mutually exclusive with the SR by an inverted
selection on \mjsum\ or \mtb. VR-\mtb\ checks the background prediction in the high-\mtb\ regime, with an upper bound on \mt\
to keep the region mutually exclusive with the corresponding SR. The other kinematic requirements are kept as close as possible to those
of the SRs to ensure that the event kinematics are similar, and allow sufficiently large yields.
 
\begin{table}[t]
\centering
\renewcommand{\arraystretch}{1.5}
\caption{Definitions of the Gtt SRs, CRs and VRs of the cut-and-count analysis.  All kinematic variables are
expressed in \gev\ except $\dphimin$, which is in radians. The jet \pt\ requirement is also applied to
$b$-tagged jets.}
\label{tab:GttEvsel}
\begin{tabular}{c c c c c c c c}
\toprule
\multicolumn{8}{c}{\textbf{ Gtt 1-lepton}}\\
\multicolumn{8}{c}{Criteria common to all regions: $\ge 1$ signal lepton, ${\pt}^\mathrm{jet} >  30~\gev$, $\nbjet \geq 3$} \\\midrule
Targeted kinematics & Type & $\njet$ & $\mt$ & $\mtb$& $\met$ & $\meffi$ & $\mjsum$ \\ \midrule
\multirow{4}{*}{\begin{minipage}{3cm}\centering Region B\\
(Boosted, Large \msplit) \end{minipage}}
& SR & $\ge 5$ & $> 150$ & $> 120 $  & $> 500 $ & $> 2200 $ & $> 200$  \\
& CR & $= 5$ & $< 150$ & $-$  & $> 300 $ & $> 1700 $ & $> 150$  \\
& VR-$\mt$ & $\ge 5$ & $> 150$ & $-$  & $> 300 $ & $> 1600 $ & $< 200$  \\
& VR-$\mtb$ & $> 5$ & $< 150$ & $> 120 $  & $> 400 $ & $> 1400 $ & $> 200$  \\\midrule
\multirow{4}{*}{\begin{minipage}{3cm}\centering Region M\\
(Moderate \msplit) \end{minipage}}
& SR & $\ge 6$ & $> 150$ & $> 160 $  & $> 450 $ & $> 1800 $ & $> 200$  \\
& CR & $= 6$ & $< 150$ & $-$  & $> 400 $ & $> 1500 $ & $> 100$  \\
& VR-$\mt$ & $\ge 6$ & $> 200$ & $-$  & $> 250 $ & $> 1200 $ & $< 100$  \\
& VR-$\mtb$ & $> 6$ & $< 150$ & $> 140 $  & $> 350 $ & $> 1200 $ & $> 150$  \\\midrule
\multirow{4}{*}{\begin{minipage}{3cm}\centering Region C\\
(Compressed, small \msplit) \end{minipage}}
& SR & $\ge 7$ & $> 150$ & $> 160 $  & $> 350 $ & $> 1000 $ & $-$  \\
& CR & $= 7$ & $< 150$ & $-$  & $> 350 $ & $> 1000 $ & $-$  \\
& VR-$\mt$ & $\ge 7$ & $> 150$ & $< 160 $  & $> 300 $ & $> 1000 $ & $-$  \\
& VR-$\mtb$ & $> 7$ & $< 150$ & $> 160 $  & $> 300 $ & $> 1000 $ & $-$  \\
\end{tabular}
\begin{tabular}{c c c c c c c c c c c}
\toprule
\multicolumn{11}{c}{\textbf{ Gtt 0-lepton}}\\
\multicolumn{11}{c}{Criteria common to all regions: ${\pt}^\mathrm{jet} > 30$~GeV} \\\midrule
Targeted kinematics & Type & $N_\mathrm{lepton}$ & $\nbjet$& $\njet$&  $\dphimin$ & $\mt$ & $\mtb$ & $\met$ & $\meffi$ & $\mjsum$ \\ \midrule
\multirow{3}{*}{\begin{minipage}{3cm}\centering Region B\\
(Boosted, Large \msplit) \end{minipage}}
& SR & $= 0$  & $\ge 3$ & $\ge 7$ & $>0.4$ & $-$ & $> 60 $ & $> 350 $ & $> 2600$ & $> 300$\\
& CR & $= 1$  & $\ge 3$ & $\ge 6$ & $-$ & $<150$ & $-$ & $> 275 $ & $> 1800$ & $> 300$\\
& VR & $= 0$  & $\ge 3$ & $\ge 6$ & $>0.4$ & $-$ & $-$ & $> 250 $ & $> 2000$ & $< 300$\\ \midrule
\multirow{3}{*}{\begin{minipage}{3cm}\centering Region M\\
(Moderate \msplit) \end{minipage}}
& SR & $= 0$  & $\ge 3$ & $\ge 7$ & $>0.4$ & $-$ & $> 120 $ & $> 500 $ & $> 1800$ & $> 200$\\
& CR & $= 1$  & $\ge 3$ & $\ge 6$ & $-$ & $<150$ & $-$ & $> 400 $ & $> 1700$ & $> 200$\\
& VR & $= 0$  & $\ge 3$ & $\ge 6$ & $>0.4$ & $-$ & $-$ & $> 450 $ & $> 1400$ & $< 200$\\ \midrule
\multirow{3}{*}{\begin{minipage}{3cm}\centering Region C\\
(Compressed, moderate \msplit) \end{minipage}}
& SR & $= 0$  & $\ge 4$ & $\ge 8$ & $>0.4$ & $-$ & $> 120 $ & $> 250 $ & $> 1000$ & $> 100$\\
& CR & $= 1$  & $\ge 4$ & $\ge 7$ & $-$ & $<150$ & $-$ & $> 250 $ & $> 1000$ & $> 100$\\
& VR & $= 0$  & $\ge 4$ & $\ge 7$ & $>0.4$ & $-$ & $-$ & $> 250 $ & $> 1000$ & $< 100$\\
 
\bottomrule
\end{tabular}
\end{table}

The signal regions of the 0-lepton channel follow a similar strategy to the 1-lepton channel.
Background composition studies performed on simulated event samples show that semileptonic \ttbar\ events, for
which the lepton is outside the acceptance or is a hadronically decaying $\tau$-lepton, dominate in the SRs.
Thus, CRs to normalise the $\ttbar$+jets background make use of the 1-lepton channel, requiring the presence of
exactly one signal lepton. An inverted selection on \mt\ is applied to remove overlaps with the 1-lepton SRs.
The background prediction is validated in a 0-lepton region, inverting the \mjsum\ selection to remove any overlap with the
SRs.

Regions targeting the Gbb model are presented in Table \ref{tab:Gbb0LEvsel}. The region definition follows the
same pattern as for Gtt-0L regions, in particular for regions B, M and
C. For very small values of \msplit, the Gbb signal does not
lead to a significant amount of \met, except if a hard ISR jet recoils against the
gluino pair. Such events are
targeted by region VC (for ``very
compressed'') that identifies an ISR-jet candidate as a non-$b$-tagged high-\pt\ leading jet (\leadjet), with a large
azimuthal separation \dphilead\ with respect to $\vec{\pt}^{\mathrm{miss}}$.
Similarly, the normalisation factor of the \ttbar\ background is
extracted from a 1-lepton CR, to which an inverted selection
on \mt\ is applied to remove the overlaps with Gtt 1-lepton SRs and the corresponding signal contamination.
The 0-lepton VRs are constructed in the 0-lepton channel with selections very close to the SR ones. They are
mutually exclusive due to an inverted \met\ selection in the VR.
 
\begin{landscape}
\begin{table}[t]
\centering
\renewcommand{\arraystretch}{1.5}
\caption{Definitions of the Gbb SRs, CRs and VRs of the cut-and-count analysis.  All kinematic variables are
expressed in \gev\ except $\dphimin$, which is in radians. The jet \pt\ requirement is applied to the
four leading jets, a subset of which are $b$-tagged jets. The $\leadjet \neq b$  requirement specifies that
the leading jet is not $b$-tagged.}
\label{tab:Gbb0LEvsel}
\begin{tabular}{c c c c c c c c c c}
\toprule
\multicolumn{10}{c}{\textbf{ Gbb}}\\
\multicolumn{10}{c}{Criteria common to all regions: $\njet \geq 4$,
${\pt}^\mathrm{jet} > 30$~GeV } \\\midrule
Targeted kinematics  & Type & $N_\mathrm{lepton}$ & $\nbjet$ &  $\dphimin$ & $\mt$ & $\mtb$ & $\met$ & $\meff$ & Others  \\\midrule
\multirow{3}{*}{\begin{minipage}{3cm}\centering Region B\\
(Boosted, Large \msplit) \end{minipage}}
& SR & $= 0$  & $\ge 3$ & $>0.4$ & $-$ & $- $ & $> 400 $ & $> 2800$ & $-$ \\
& CR & $= 1$  & $\ge 3$ & $-$ & $< 150$ & $- $ & $> 400 $ & $> 2500$ & $-$ \\
& VR & $= 0$  & $\ge 3$ & $>0.4$ & $-$ & $- $ & $> 350 $ & $1900$--$2800$ & $-$ \\\midrule
\multirow{3}{*}{\begin{minipage}{3cm}\centering Region M\\
(Moderate \msplit) \end{minipage}}
& SR & $= 0$  & $\ge 4$ & $>0.4$ & $-$ & $>90$ & $> 450 $ & $> 1600$ & $-$ \\
& CR & $= 1$  & $\ge 4$ & $-$ & $< 150$ & $- $ & $> 300 $ & $> 1600$ & $-$ \\
& VR & $= 0$  & $\ge 4$ & $>0.4$ & $-$ & $>100$ & $250$--$450$ & $1600$--$1900$ & $-$ \\\midrule
\multirow{3}{*}{\begin{minipage}{3cm}\centering Region C\\
(Compressed, small \msplit) \end{minipage}}
& SR & $= 0$  & $\ge 4$ & $>0.4$ & $-$ & $>155$ & $> 450 $ & $-$ & $-$ \\
& CR & $= 1$  & $\ge 4$ & $-$ & $< 150$ & $- $ & $> 375 $ & $-$ & $-$ \\
& VR & $= 0$  & $\ge 4$ & $>0.4$ & $-$ & $>125$ & $350$--$450$ & $-$ & $-$ \\\midrule
\multirow{3}{*}{\begin{minipage}{3cm}\centering Region VC\\
(Very Compressed, very small \msplit) \end{minipage}}
& SR & $= 0$  & $\ge 3$ & $>0.4$ & $-$ & $>100$ & $> 600 $ & $-$ &
\multirow{3}{*}{\begin{minipage}{3cm}\centering $\pt^{\leadjet}>400$, $\leadjet \neq b$, $\dphilead>2.5$\end{minipage}} \\
& CR & $= 1$  & $\ge 3$ & $-$ & $< 150$ & $- $ & $> 600 $ & $-$ \\
& VR & $= 0$  & $\ge 3$ & $>0.4$ & $-$ & $>100$ & $225$--$600$ & $-$ \\
\bottomrule
\end{tabular}
\end{table}
\end{landscape}

\subsection{Multi-bin analysis}
\label{sec:multibin}

Figures \ref{fig:presel_0l} and \ref{fig:presel_1l} show that a good separation between signal and background
can be achieved with various kinematic variables. The distribution of $\njet$ and \meff\ for different signal
benchmarks and \msplit\ values is used to build a
two-dimensional slicing of the phase space in a set of non-overlapping SRs, CRs and VRs
that can be statistically combined. The slicing scheme is presented in Figure \ref{fig:multibin_scheme}.
The SRs are named in the form SR-{\textit Y}L-{\textit Z$_1$Z$_2$}, where
{\textit Y} indicates the number of leptons,  {\textit Z$_1$} labels
the jet multiplicity bin and {\textit Z$_2$} labels the \meff\
bin. For {\textit Z$_1$} and {\textit Z$_2$}, the letters ``H''
stands for ``high'', ``I'' for ``intermediate'' and ``L'' for
``low''. In the 0-lepton channel, there is also a 0L-ISR
region that is a subset of the IL, LL, II and LI regions, and kept mutually exclusive with them as detailed below.
 
\begin{figure}[b]
\subfigure[]{\includegraphics[width=0.49\linewidth]{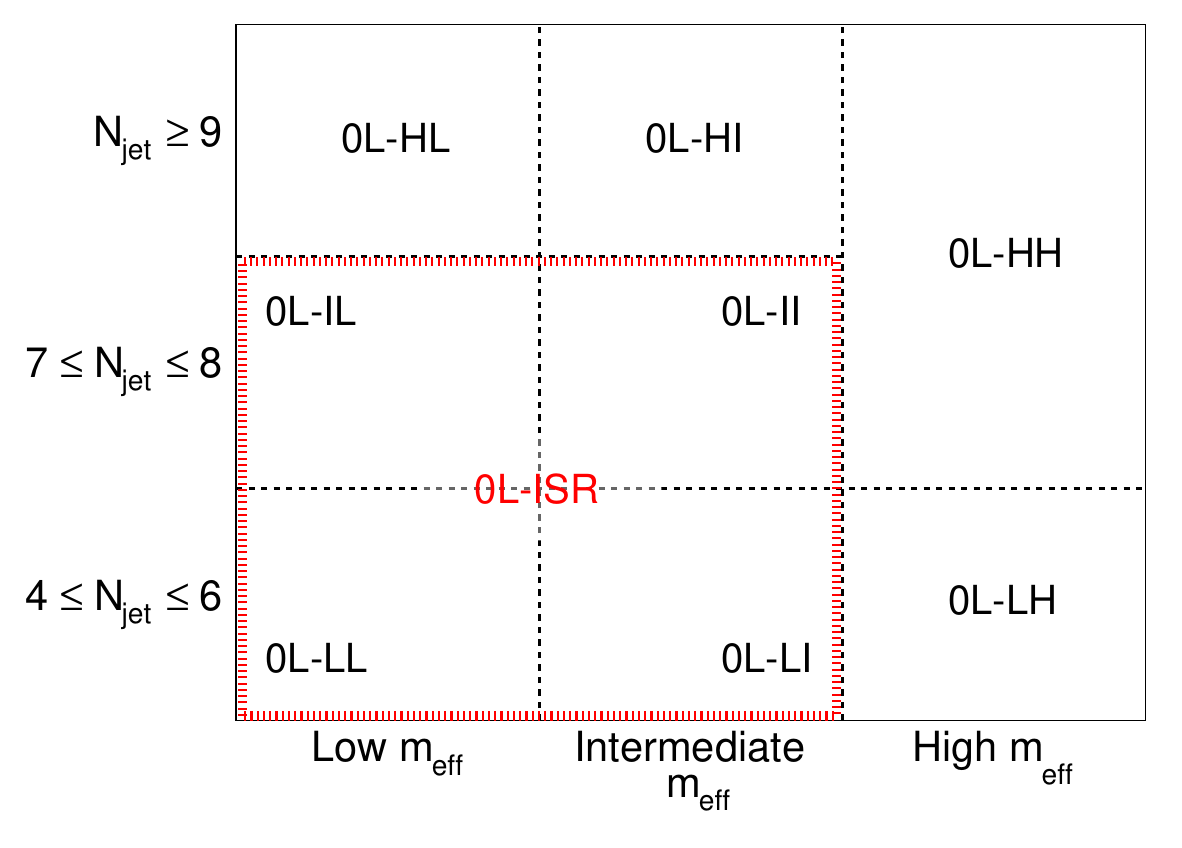}\label{fig:multibin_scheme_0l}}
\subfigure[]{\includegraphics[width=0.49\linewidth]{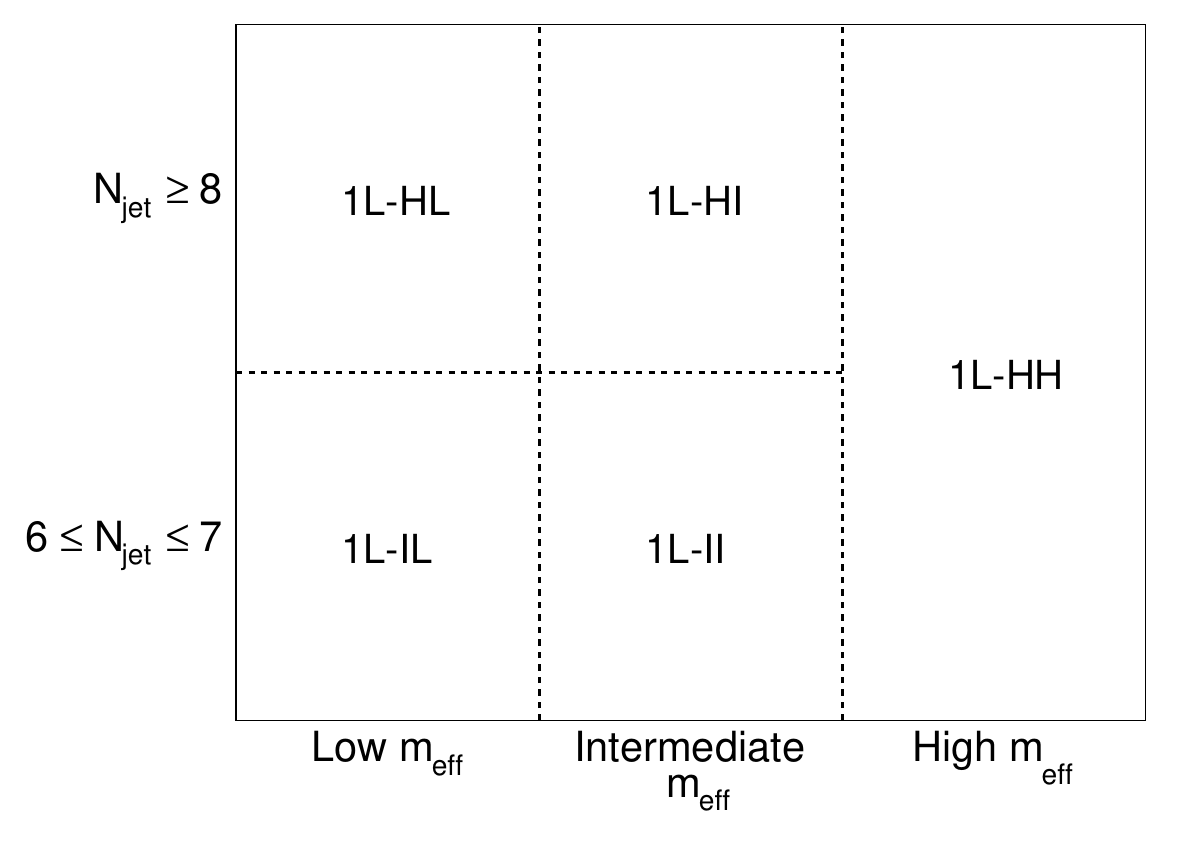}\label{fig:multibin_scheme_1l}}
\caption{Scheme of the multi-bin analysis for the \subref{fig:multibin_scheme_0l} 0-lepton
and \subref{fig:multibin_scheme_1l} 1-lepton regions.
The 0L-ISR region is represented with the broad red
dashed line in \subref{fig:multibin_scheme_0l}.
}
\label{fig:multibin_scheme}
\end{figure}
 
The low-$\njet$ region probes especially Gbb-like models, for which the number of hard jets is lower than in decay
topologies containing top quarks. This category of events is thus only considered in the 0-lepton channel. Gtt
events are mostly expected in the high-$\njet$ bin. The intermediate jet multiplicity bin is built to be sensitive to
decay topologies with a number of top quarks intermediate between Gbb and Gtt, but also to Gbb (with additional
jets originating from radiation) and to Gtt (when some jets fall outside the acceptance). The \meff\ bins are chosen
to provide sensitivity to various kinematic regimes: the low-\meff\ regions are essentially sensitive to soft signals (low
\msplit), while the high-\meff\ regions are designed to select highly boosted events.
 
For each $\njet$--\meff\ region presented in Figure
\ref{fig:multibin_scheme}, the selection was optimised over
all the other variables to maximise the exclusion power for the
Gbb and Gtt models. For each \meff\ bin, a targeted range of \msplit\ was used in the optimisation
procedure.
 
The high- and intermediate-$\njet$ regions are presented in Tables
\ref{tab:multibin_Hn} and \ref{tab:multibin_In}, respectively.
For each \meff\ region, 0- and 1-lepton channels are used to
provide sensitivity to the Gtt model and the decay topologies of the
variable branching ratio model which contain at least one top quark.
In the intermediate-$\njet$ categories the leading jet is required to be
$b$-tagged or the value
of $\dphilead$ to be lower than 2.9 in order to ensure they are mutually exclusive with
the 0L-ISR regions.
Corresponding 0-lepton and 1-lepton SRs
share a single CR, hosted in the 1-lepton channel, after the application of an inverted
\mt\ selection to remove the overlap with the 1-lepton SRs. The other kinematic
requirements are kept close to the ones of the SR. One VR is defined for
each SR in the corresponding lepton channel. Full independence between the signal and
VRs is guaranteed by \met\ and \mtb\ requirements.
 
The low-$\njet$ regions are presented in Table \ref{tab:multibin_Ln}. Targeting primarily the Gbb model,
the transverse momentum of the fourth jet is required to be larger than 90~\gev\
in all SRs. In the intermediate and low \meff\ regions, the leading jet is required to be $b$-tagged
or the value of $\dphilead$ to be lower than 2.9 in order to be mutually exclusive with the 0L-ISR regions. The \ttbar\
background dominates in all regions, and is normalised in dedicated
1-lepton regions, defined with a low \mt\ requirement, as done for the regions of the cut-and-count analysis. VRs
are constructed in the 0-lepton channel, closely reproducing the background composition and kinematics of the
SR events.
 
A dedicated set of regions is designed to target very compressed Gbb scenarios in which a hard ISR jet
recoils against the gluino pair. The definition of these regions is presented in Table \ref{tab:multibin_ISR}.

\begin{landscape}
\begin{table}[t]
\centering
\renewcommand{\arraystretch}{1.5}
\caption{Definition of the high-$\njet$ SRs, CRs and VRs of the multi-bin analysis. All kinematic variables are
expressed in \gev\ except $\dphimin$, which is in radians.}
\label{tab:multibin_Hn}
\begin{tabular}{c c c c c c c c c c}
\toprule
\multicolumn{10}{c}{\textbf{ High-$\njet$ regions}}\\
\multicolumn{10}{c}{Criteria common to all regions: $\nbjet \geq 3$, ${\pt}^\mathrm{jet} > 30$~GeV } \\
\midrule
Targeted kinematics  & Type & $N_\mathrm{lepton}$ & $\dphimin$ & $\mt$ & $\njet$ & $\mtb$ & $\mjsum$ & $\met$ & $\meff$  \\
\midrule
\multirow{5}{*}{\begin{minipage}{3cm}\centering High-\meff\ \\ (HH) \\ (Large \msplit) \end{minipage}}
& SR-0L 	& $= 0$  		& $>0.4$ 		& $-$ 		& $\ge 7$		& $>100 $ 			& $>200$ 	& $> 400 $ 				& $> 2500$ \\
& SR-1L 	& $\ge 1$  	& $-$		& $> 150 $ 	& $\ge 6$		& $> 120$ 			& $>200$ 	& $> 500 $ 				& $> 2300$ \\
& CR 	& $\ge 1$  	& $-$ 		& $< 150$ 	& $\ge 6$		& $> 60 $ 				& $>150$ 	& $> 300 $ 				& $> 2100$ \\
& VR-0L 	& $= 0$  		& $>0.4$ 		& $-$ 		& $\ge 7$		& $<100$ if $\met>300$ 	& $-$ 	& $< 300 $ if $\mtb > 100$ 	& $> 2100$ \\
& VR-1L 	& $\ge 1$  	& $-$ 		& $> 150$ 	& $\ge 6$		& $<140$ if $\meff>2300$	& $-$ 	& $< 500$  				& $> 2100$ \\
\midrule
\multirow{5}{*}{\begin{minipage}{3cm}\centering Intermediate-\meff\ \\ (HI) \\ (Intermediate \msplit) \end{minipage}}
& SR-0L 	& $= 0$  		& $>0.4$ 		& $-$ 		& $\ge 9$		& $> 140$ 			& $>150$ 	& $> 300 $ 				& $[1800, 2500]$ \\
& SR-1L 	& $\ge 1$  	& $-$		& $> 150 $ 	& $\ge 8$		& $> 140$ 			& $>150$ 	& $> 300 $ 				& $[1800, 2300]$ \\
& CR 	& $\ge 1$  	& $-$ 		& $< 150$ 	& $\ge 8$		& $> 60$ 				& $>150$ 	& $> 200 $ 				& $[1700, 2100]$ \\
& VR-0L 	& $= 0$  		& $>0.4$ 		& $-$ 		& $\ge 9$		& $<140$ if $\met>300$ 	& $-$ 	& $< 300 $ if $\mtb > 140$ 	& $[1650, 2100]$ \\
& VR-1L 	& $\ge 1$  	& $-$ 		& $> 150$ 	& $\ge 8$		& $<140$ if $\met>300$	& $-$ 	& $< 300 $ if $\mtb > 140$	& $[1600, 2100]$ \\
\midrule
\multirow{5}{*}{\begin{minipage}{3cm}\centering Low-\meff\ \\ (HL) \\ (Small \msplit) \end{minipage}}
& SR-0L 	& $= 0$  		& $>0.4$ 		& $-$ 		& $\ge 9$		& $> 140$ 			& $-$ 	& $> 300 $ 				& $[900, 1800]$ \\
& SR-1L 	& $\ge 1$  	& $-$		& $> 150 $ 	& $\ge 8$		& $> 140$ 			& $-$ 	& $> 300 $ 				& $[900, 1800]$ \\
& CR 	& $\ge 1$  	& $-$ 		& $< 150$ 	& $\ge 8$		& $> 130$ 			& $-$ 	& $> 250 $ 				& $[900, 1700]$ \\
& VR-0L 	& $= 0$  		& $>0.4$ 		& $-$ 		& $\ge 9$		& $<140$				& $-$ 	& $> 300 $ 				& $[900, 1650]$ \\
& VR-1L 	& $\ge 1$  	& $-$ 		& $> 150$ 	& $\ge 8$		& $<140$				& $-$ 	& $> 225 $			 	& $[900, 1650]$ \\
\bottomrule
\end{tabular}
\end{table}
\end{landscape}
 
\begin{landscape}
\begin{table}[t]
\small
\centering
\renewcommand{\arraystretch}{1.5}
\caption{Definition of the intermediate-$\njet$ SRs, CRs and VRs of the multi-bin analysis. All kinematic variables are
expressed in \gev\ except $\dphimin$, which is in radians. The $\leadjet = b$  requirement specifies that
the leading jet is $b$-tagged.}
\label{tab:multibin_In}
\begin{tabular}{c c c c c c c c c c c}
\toprule
\multicolumn{11}{c}{\textbf{ Intermediate-$\njet$ regions}}\\
\multicolumn{11}{c}{Criteria common to all regions: $\nbjet \geq 3$, ${\pt}^\mathrm{jet} > 30$~GeV } \\
\midrule
Targeted kinematics  & Type & $N_\mathrm{lepton}$ & $\dphimin$ & $\mt$ & $\njet$ & $\leadjet = b$ or $\dphilead \leq 2.9$ & $\mtb$ & $\mjsum$  & $\met$ & $\meff$  \\
\midrule
\multirow{5}{*}{\begin{minipage}{3cm}\centering Intermediate-\meff\ \\ (II) \\ (Intermediate \msplit) \end{minipage}}
& SR-0L 	& $= 0$  		& $>0.4$ 		& $-$ 		& $[7,8]$		& \cmark		& $> 140 $ 			& $>150$ 		& $> 300 $ 				& $[1600,2500]$ \\
& SR-1L 	& $\ge 1$  	& $-$		& $> 150 $ 	& $[6,7]$		& $-$ 		& $> 140$ 			& $>150$ 		& $> 300 $ 				& $[1600,2300]$ \\
& CR 	& $\ge 1$  	& $-$ 		& $< 150$ 	& $[6,7]$		& \cmark 		& $> 110 $ 			& $>150$ 		& $> 200 $ 				& $[1600,2100]$ \\
& VR-0L 	& $= 0$  		& $>0.4$ 		& $-$ 		& $[7,8]$		& \cmark		& $<140$ 				& $-$ 		& $> 300 $ 				& $[1450,2000]$ \\
& VR-1L 	& $\ge 1$  	& $-$ 		& $> 150$ 	& $[6,7]$		& $-$		& $<140$				& $-$ 		& $> 225 $  				& $[1450,2000]$ \\
\midrule
\multirow{5}{*}{\begin{minipage}{3cm}\centering Low-\meff\ \\ (IL) \\ (Low \msplit) \end{minipage}}
& SR-0L 	& $= 0$  		& $>0.4$ 		& $-$ 		& $[7,8]$		& \cmark		& $> 140 $ 			& $-$ 		& $> 300 $ 				& $[800,1600]$ \\
& SR-1L 	& $\ge 1$  	& $-$		& $> 150 $ 	& $[6,7]$		& $-$ 		& $> 140$ 			& $-$ 		& $> 300 $ 				& $[800,1600]$ \\
& CR 	& $\ge 1$  	& $-$ 		& $< 150$ 	& $[6,7]$		& \cmark 		& $> 130 $ 			& $-$ 		& $> 300 $ 				& $[800,1600]$ \\
& VR-0L 	& $= 0$  		& $>0.4$ 		& $-$ 		& $[7,8]$		& \cmark		& $<140$ 				& $-$ 		& $> 300 $ 				& $[800,1450]$ \\
& VR-1L 	& $\ge 1$  	& $-$ 		& $> 150$ 	& $[6,7]$		& $-$		& $<140$				& $-$ 		& $> 300 $  				& $[800,1450]$ \\
\bottomrule
\end{tabular}
\end{table}
\end{landscape}

\begin{landscape}
\begin{table}[t]
\centering
\renewcommand{\arraystretch}{1.5}
\caption{Definition of the low-$\njet$ and ISR SRs, CRs and VRs of the multi-bin analysis. All kinematic variables are
expressed in \gev\ except $\dphimin$, which is
in radians. The $\leadjet = b$ ($\leadjet \neq b$) requirement specifies that
the leading jet is (not) $b$-tagged.}
\label{tab:multibin_Ln}
\label{tab:multibin_ISR}
\begin{tabular}{c c c c c c c c c c c }
\toprule
\multicolumn{11}{c}{\textbf{ Low-$\njet$ regions}}\\
\multicolumn{11}{c}{Criteria common to all regions: $\nbjet \geq 3$, ${\pt}^\mathrm{jet} > 30$~GeV } \\
\midrule
Targeted kinematics  & Type & $N_\mathrm{lepton}$ & $\dphimin$ & $\mt$ & $\njet$ & $\leadjet = b$ or $\dphilead \leq 2.9$ & $\pt^{\fourthjet}$ & $\mtb$ & $\met$ & $\meff$  \\
\midrule
\multirow{3}{*}{\begin{minipage}{3cm}\centering High-\meff\ \\ (LH) \\ (Large \msplit) \end{minipage}}
& SR 	& $= 0$  		& $>0.4$ 		& $-$ 		& $[4,6]$		& $-$		& $>90$					& $-$ 			& $> 300 $ 				& $> 2400$ \\
& CR 	& $\ge 1$  	& $-$ 		& $< 150$ 	& $[4,5]$		& $-$ 	 	& -						& $-$ 			& $> 200 $ 				& $> 2100$ \\
& VR 	& $= 0$  		& $>0.4$ 		& $-$ 		& $[4,6]$		& $-$ 		& $>90$ if $\met < 300$		& $-$			& $> 200 $ 				& $[2000,2400]$ \\
\midrule
\multirow{3}{*}{\begin{minipage}{3cm}\centering Intermediate-\meff\ \\ (LI) \\ (Intermediate \msplit) \end{minipage}}
& SR 	& $= 0$  		& $>0.4$ 		& $-$ 		& $[4,6]$		& \cmark		& $>90$				& $>140$ 		& $> 350 $ 				& $[1400,2400]$ \\
& CR 	& $\ge 1$  	& $-$ 		& $< 150$ 	& $[4,5]$		& \cmark 	 	& $>70$				& $-$ 		& $> 300 $ 				& $[1400,2000]$ \\
& VR 	& $= 0$  		& $>0.4$ 		& $-$ 		& $[4,6]$		& \cmark		& $>90$				& $<140$ 		& $> 300 $ 				& $[1250,1800]$ \\
\midrule
\multirow{3}{*}{\begin{minipage}{3cm}\centering Low-\meff\ \\ (LL) \\ (Low \msplit) \end{minipage}}
& SR 	& $= 0$  		& $>0.4$ 		& $-$ 		& $[4,6]$		& \cmark		& $>90$				& $>140$ 		& $> 350 $ 				& $[800,1400]$ \\
& CR 	& $\ge 1$  	& $-$ 		& $< 150$ 	& $[4,5]$		& \cmark 	 	& $>70$				& $-$ 		& $> 300 $ 				& $[800,1400]$ \\
& VR 	& $= 0$  		& $>0.4$ 		& $-$ 		& $[4,6]$		& \cmark		& $>90$				& $<140$ 		& $> 300 $ 				& $[800,1250]$ \\
\bottomrule
\end{tabular}
 
\vspace{1cm}
 
\begin{tabular}{K{1.5cm} K{1.5cm} K{1.5cm} K{1.5cm} K{1.5cm} K{1.5cm} K{1.5cm} K{1.5cm} }
\toprule
\multicolumn{8}{c}{\textbf{ ISR regions}}\\
\multicolumn{8}{c}{Criteria common to all regions: $\nbjet \geq 3$, $\dphilead > 2.9$, ${\pt}^\leadjet > 400$~\gev, ${\pt}^\mathrm{jet} > 30$~GeV, $\leadjet \neq b$} \\
\midrule
Type & $N_\mathrm{lepton}$ & $\dphimin$ & $\mt$ & $\njet$ & $\mtb$ & $\met$ & $\meff$  \\
\midrule
SR 	& $= 0$  		& $>0.4$ 		& $-$ 		& $[4,8]$		& $>100$ 			& $> 600 $ 				& $<2200$ \\
CR 	& $\ge 1$  	& $-$ 		& $< 150$ 	& $[4,7]$		& $-$ 			& $> 400 $ 				& $<2000$ \\
VR 	& $= 0$  		& $>0.4$ 		& $-$ 		& $[4,8]$		& $>100$ 			& $> 250 $ 				& $<2000$ \\
\bottomrule
\end{tabular}
\end{table}
\end{landscape}

\FloatBarrier

 
\section{Systematic uncertainties}
\label{sec:syst}
Figures \ref{fig:syst_cutandcount} and \ref{fig:syst_multibin} summarise the relative systematic uncertainties in the background estimate for the cut-and-count and multi-bin analyses, respectively. These uncertainties arise from the extrapolation of the $\ttbar$ normalisation obtained in the CRs to the SRs as well as from the yields of the minor backgrounds in the SRs, which are predicted by the simulation.
The total systematic uncertainties range from approximately 20\% to 80\% in the various SRs.
 
\begin{figure}[htbp]
\centering
\subfigure[]{\includegraphics[width=0.85\textwidth]{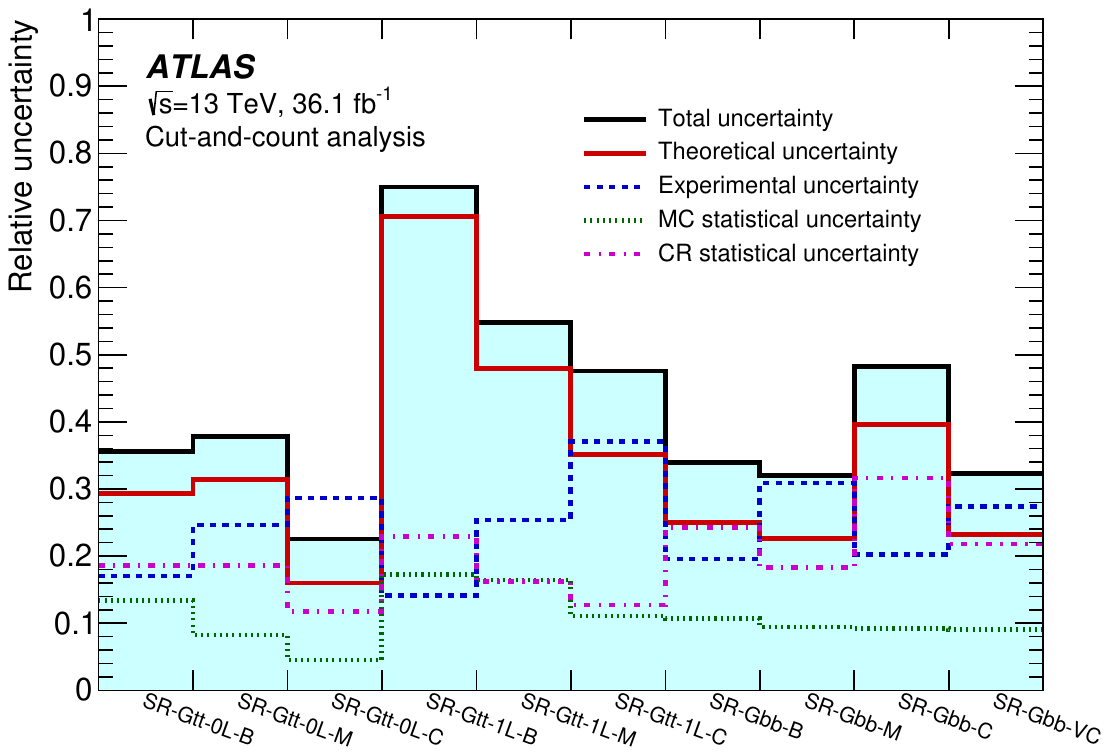}\label{fig:syst_cutandcount}}\\
\subfigure[]{\includegraphics[width=0.85\textwidth]{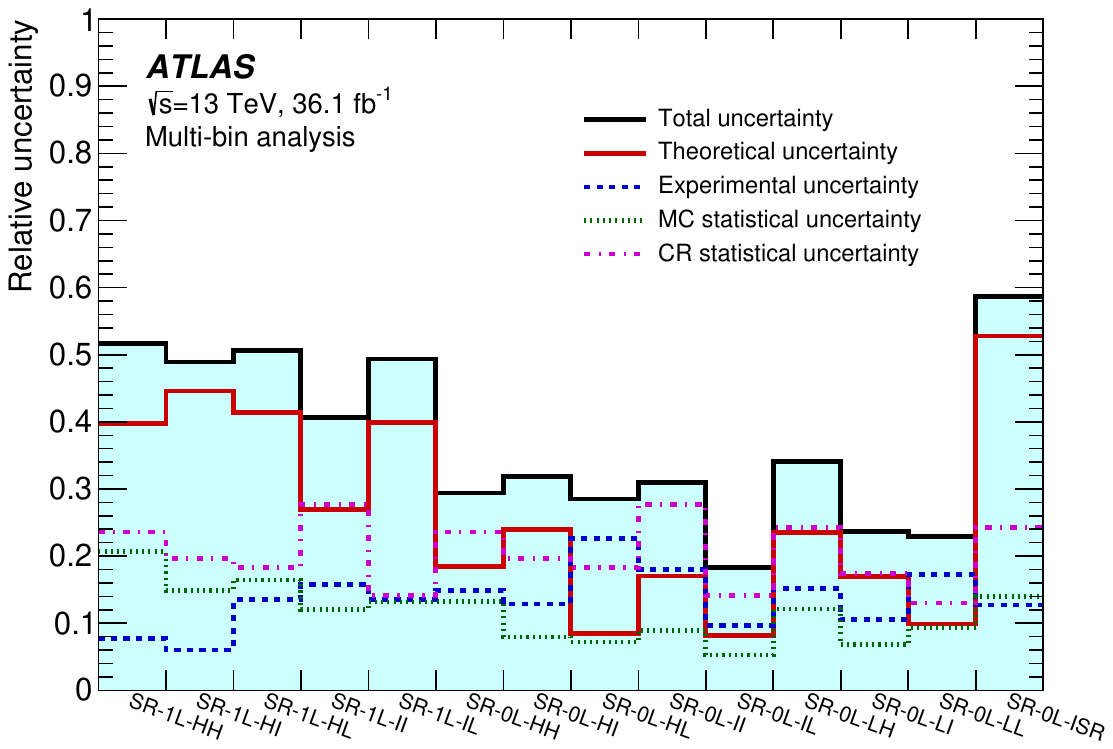}\label{fig:syst_multibin}}\\
\caption{Relative systematic uncertainty in the background estimate for the \subref{fig:syst_cutandcount} cut-and-count and \subref{fig:syst_multibin} multi-bin analyses. The individual uncertainties can be correlated, such that the total background uncertainty is not necessarily their sum in quadrature.
}
\label{fig:syst}
\end{figure}
 
The detector-related systematic uncertainties affect both the background estimate and the signal yield.
The largest sources in this analysis relate to
the jet energy scale, jet energy resolution (JER) and the $b$-tagging efficiencies and
mistagging rates. The JES uncertainties for the small-$R$ jets are derived from $\sqrt{s}= 13~\tev$
data and simulations while the JER uncertainties are extrapolated
from 8~\tev\ data using MC simulations~\cite{ATL-PHYS-PUB-2015-015}. These uncertainties are also propagated to the re-clustered
large-$R$ jets, which use them as inputs. The jet mass scale and resolution uncertainties have a negligible impact on
the re-clustered jet mass. The impact of the JES uncertainties on the expected background yields is
between 4\% and 35\%, while JER uncertainties affect the background yields by approximately 0--26\% in the
various regions. Uncertainties in the measured $b$-tagging efficiencies and mistagging rates are the subleading sources
of experimental uncertainty.
 
The impact of these uncertainties on the expected background
yields is 3--24\% depending on the considered region.
The uncertainties associated with lepton reconstruction and energy measurements have a negligible impact
on the final results. All lepton and jet measurement uncertainties are propagated to the calculation
of \met, and additional uncertainties are included in the scale and resolution of the soft term. The
overall impact of the \met soft-term uncertainties is also small.

Since the normalisation of the \ttbar background is fit to data in the CRs, uncertainties in the
modelling of this background only affect the extrapolation from the CRs to the SRs and VRs.
Hadronisation and parton showering model uncertainties are estimated using
a sample generated with \POWHEG  and showered by \MYHERWIG++ v2.7.1 with the UEEE5
underlying-event tune.
Systematic uncertainties in the modelling of  initial- and final-state radiation are explored with \POWHEG samples
showered with two alternative settings of \PYTHIA v6.428.
The first of these uses the PERUGIA2012radHi tune~\cite{Skands:2010ak} and has the renormalisation
and factorisation scales set to twice the nominal value, resulting in more radiation in the final state.
In addition, it has $h_\mathrm{damp}$ set to 2$m_\mathrm{top}$. The second sample, using the PERUGIA2012radLo tune,
has $h_\mathrm{damp}=m_\mathrm{top}$ and the renormalisation and factorisation scales are set to half of
their nominal values, resulting in less radiation in the event. In each case, the uncertainty is taken as
the change in the expected yield of \ttbar background with respect to the nominal sample. The uncertainty
due to the choice of event generator is estimated by comparing the expected yields obtained using a \ttbar sample
generated with \MGMCatNLO and one that is generated with \POWHEG. Both of these samples are showered with
\MYHERWIG++ v2.7.1. The total theoretical uncertainty in the inclusive \ttbar background is taken as the sum in quadrature of these
individual components. An additional uncertainty is assigned to the fraction of \ttbar events produced in association with additional
heavy-flavour jets (i.e. $\ttbar + \geq \mathrm{1}b$ and $\ttbar + \geq \mathrm{1}c$), a process which suffers from large theoretical uncertainties.
Simulation studies show that the heavy-flavour fractions in each set of SR, CR and VR, which have almost identical
$b$-tagged jets requirements, are similar.  Therefore, the theoretical uncertainties in this fraction affect these regions in
a similar way, and thus largely cancel out in the semi-data-driven $\ttbar$ normalisation based on the observed CR yields.
The residual uncertainty in the \ttbar prediction is taken as the difference between the nominal \ttbar prediction and
the one obtained after varying the cross-section of \ttbar events with
additional heavy-flavour jets by 30\%, in accordance with the results of the ATLAS measurement of this cross-section
at $\sqrt{s}= 8~\tev$~\cite{TOPQ-2014-10}. This component typically makes a small contribution (0--8\%) to the total impact of the \ttbar\ modelling
uncertainties on the background yields, which ranges between 5\% and 76\% for the various regions.
The statistical uncertainty of the CRs used to extract the $\ttbar$ normalisation factors, which is included in the systematic uncertainties, ranges from 10\% to 30\% depending on the SR.

Modelling uncertainties affecting the single-top process arise especially from the interference between the \ttbar\ and $Wt$ processes.
This uncertainty is estimated using inclusive $WWbb$ events, generated using \MGMCatNLO, which are compared with the sum of \ttbar\
and $Wt$ processes. Furthermore, as in the \ttbar\ modelling uncertainties, variations of \PYTHIA v6.428 settings increasing or decreasing the
amount of radiation are also used. An additional 5\% uncertainty is included in the cross-section of
single-top processes~\cite{Kant:2014oha}.
Overall, the modelling uncertainties affecting the single-top process lead to changes of approximately 0--11\% in the total
yields in the various regions.
Uncertainties in the $W/Z$+jets backgrounds are estimated by varying independently the scales for factorisation, renormalisation and resummation
by factors of 0.5 and 2. The scale used for the matching between jets originating from the matrix element and the parton shower is also varied. The resulting
uncertainties in the total yield range from approximately 0 to 50\% in the various regions.
A 50\% normalisation uncertainty is assigned to $\ttbar+W/Z/h$, $\ttbar\ttbar$ and diboson backgrounds and are found to have no significant impact
on the sensitivity of this analysis.
Uncertainties arising from variations of the parton distribution functions were found to affect background yields by less than 2\%, and
therefore these uncertainties are neglected here. Uncertainties due to the limited number of events in the MC background samples are included if above 5\%.
They reach approximately
20\% in regions targeting large mass-splitting.
 
The uncertainties in the cross-sections of signal processes are determined from an envelope of different
cross-section predictions, as described in Section~\ref{sec:samples}.
A systematic uncertainty is also assigned to the kinematic correction described in Section~\ref{sec:presel}. The total size of the
correction is used as an uncertainty, and is applied to all simulated event samples for the 1-lepton channel.
 
\section{Results}
\label{sec:results}
The expected SM background is determined separately in each SR with
a profile likelihood fit \cite{Cowan:2010js} implemented in the
HistFitter framework \cite{HFpaper}, referred to as a background-only fit. The fit uses
as a constraint the observed event yield in the associated CR to adjust the
$\ttbar$ normalisation, assuming that no signal contributes to this yield, and applies
that normalisation factor to the number of $\ttbar$ events predicted by simulation in
the SR. The values of the normalisation factors, the expected numbers of background events and
the observed data yields in all the CRs are shown in Figures \ref{fig:pullCR_discovery} and
\ref{fig:pullCR_exclusion} for the cut-and-count and multi-bin analyses, respectively.
 
\begin{figure}[htbp]
\centering
\subfigure[]{\includegraphics[width=0.9\textwidth]{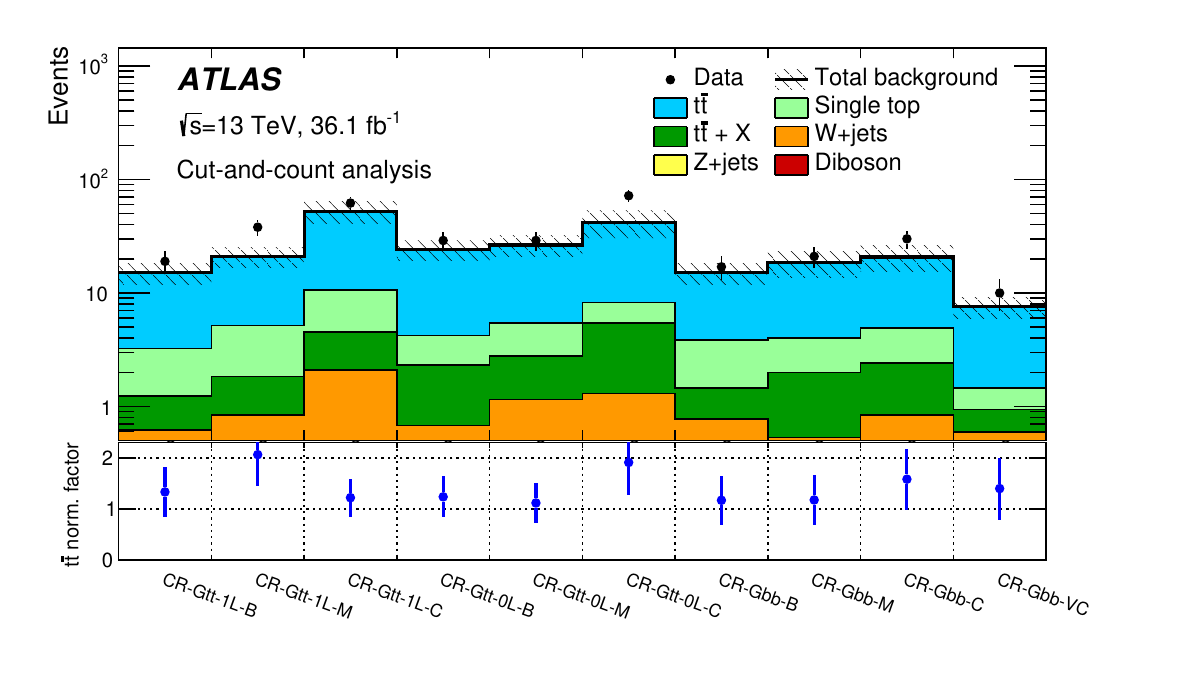}\label{fig:pullCR_discovery}}\\
\subfigure[]{\includegraphics[width=0.9\textwidth]{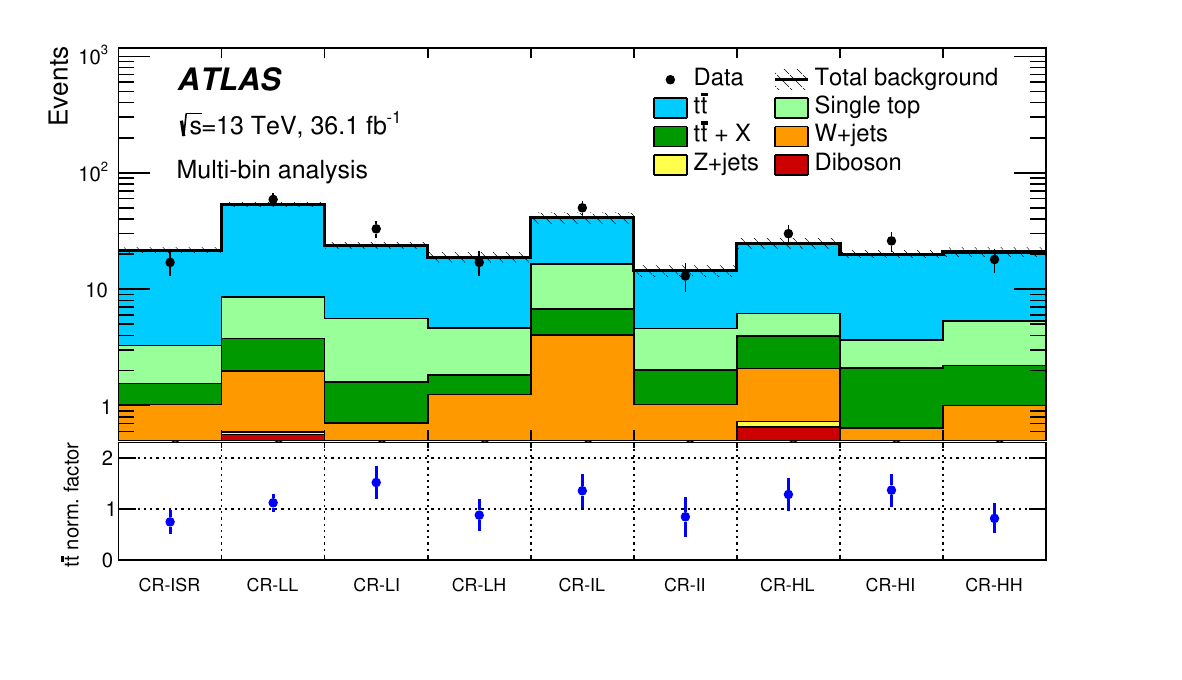}\label{fig:pullCR_exclusion}}\\
\caption{Pre-fit event yield in control regions and related \ttbar\
normalization factors after the background-only fit for
\subref{fig:pullCR_discovery}
the cut-and-count and \subref{fig:pullCR_exclusion} the multi-bin analyses. The upper panel shows
the observed number of events and the predicted background yield before the fit.
The background category $\ttbar+X$ includes $\ttbar W/Z$, $\ttbar H$ and $\ttbar \ttbar$ events. All of these
regions require at least one signal lepton, for which the
multijet background is negligible. All uncertainties describes in Section \ref{sec:syst} are included in the uncertainty band.
The $\ttbar$ normalisation is obtained from the fit
and is displayed in the bottom panel.
}
\label{fig:pullCR}
\end{figure}
 
The inputs to the background-only fit for each SR are the number of events
observed in its associated CR and the number of events predicted by
simulation in each region for all background processes. The numbers of observed and
predicted events in each CR are described by Poisson probability density
functions. The systematic uncertainties in the expected values are included in the fit
as nuisance parameters. They are constrained by Gaussian distributions with widths
corresponding to the sizes of the uncertainties and are treated as correlated, when
appropriate, between the various regions. The product of the various probability density
functions forms the likelihood, which the fit maximises by adjusting the $\ttbar$
normalisation and the nuisance parameters.
 
Figures \ref{fig:pullVR_discovery} and \ref{fig:pullVR_exclusion} show the results of the
background-only fit to the CRs, extrapolated to the VRs for the
cut-and-count and multi-bin analyses, respectively.
The number of events predicted by the background-only fit is compared to the data in
the upper panel. The pull, defined by the difference between the observed number of
events ($n_\mathrm{obs}$) and the predicted background yield ($n_\mathrm{pred}$)
divided by the total uncertainty ($\sigma_\mathrm{tot}$), is shown for each region in the
lower panel. No evidence of significant background mismodelling is observed in the
VRs.
 
\begin{figure}[htbp]
\centering
\subfigure[]{\includegraphics[width=0.9\textwidth]{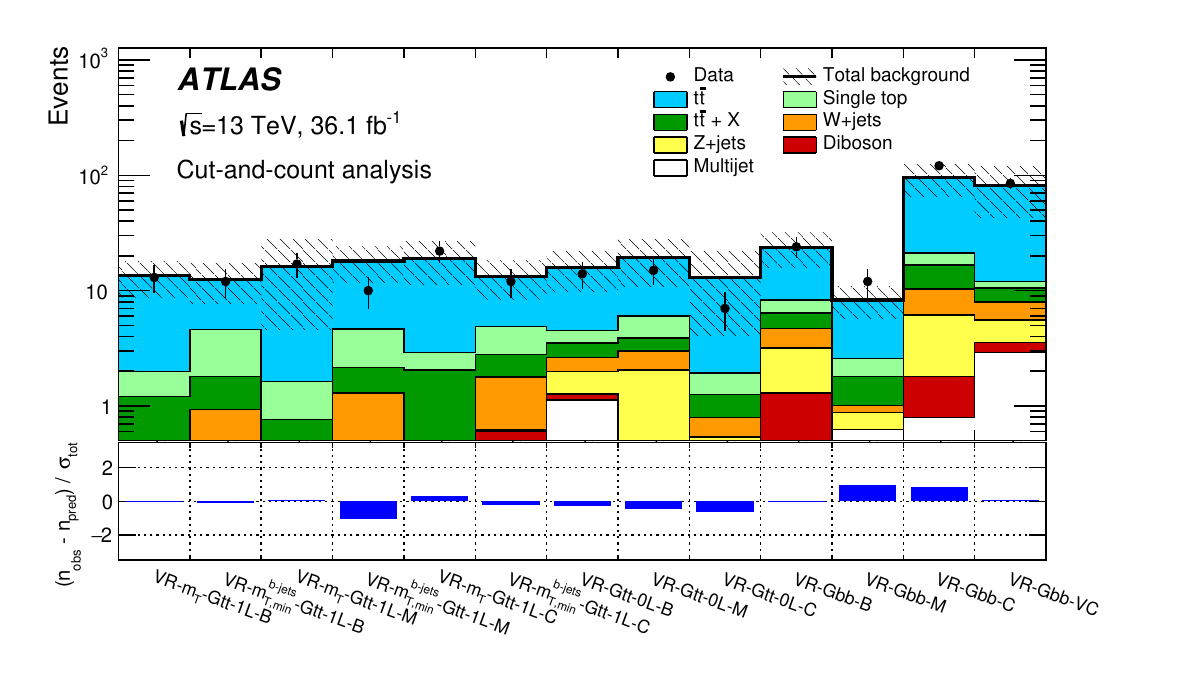}\label{fig:pullVR_discovery}}\\
\subfigure[]{\includegraphics[width=0.9\textwidth]{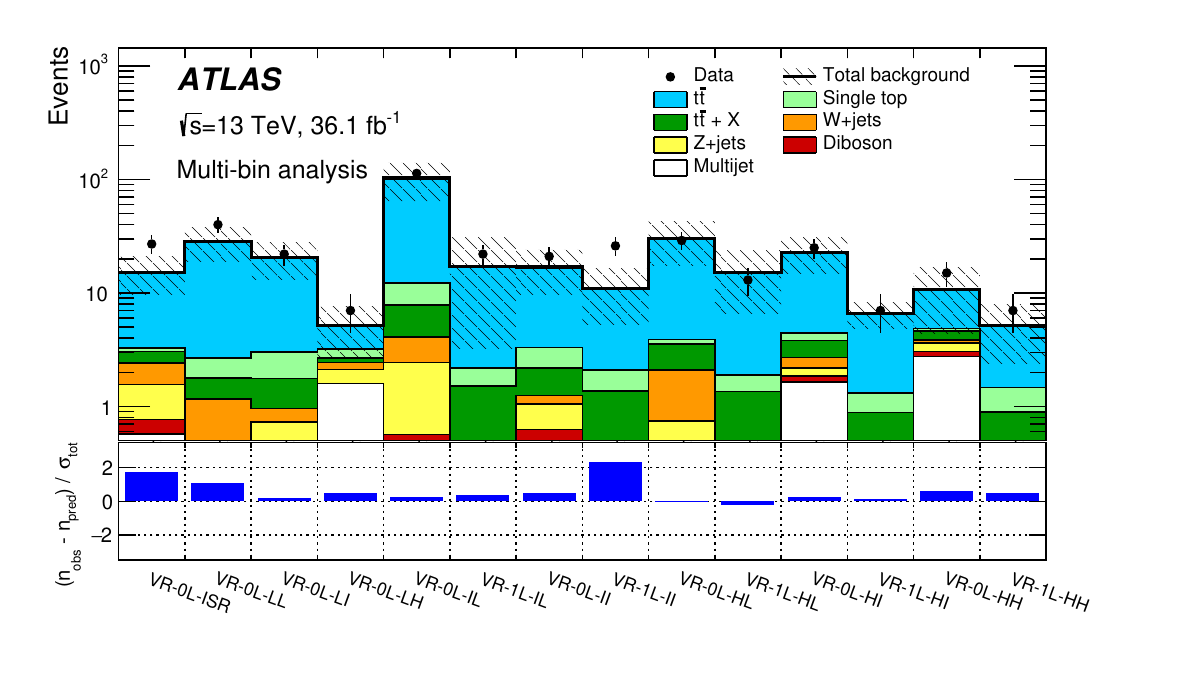}\label{fig:pullVR_exclusion}}\\
\caption{Results of the background-only fit extrapolated to the VRs of \subref{fig:pullVR_discovery} the cut-and-count and \subref{fig:pullVR_exclusion}
the multi-bin analyses. The $\ttbar$ normalisation
is obtained from the fit to the CRs shown in Figure~\ref{fig:pullCR}. The upper panel shows
the observed number of events and the predicted background yield.
All uncertainties  defined in Section~\ref{sec:syst} are included in the
uncertainty band. The background category $\ttbar+X$ includes $\ttbar W/Z$,
$\ttbar H$ and $\ttbar \ttbar$ events. The lower panel shows the pulls in
each VR.
}
\label{fig:pullVR}
\end{figure}
 
The event yields in the SRs for the cut-and-count and multi-bin analyses are presented in Figure~\ref{fig:pullSR}, where
the pull is shown for each region in the lower panel. No significant excess is found above the predicted background. The
maximum deviation is observed in region SR-0L-HH of the multi-bin analysis with a local significance of 2.3 standard
deviations. The background
is dominated by $\ttbar$ events in all SRs. The subdominant background contributions in the 0-lepton regions are
$Z(\to \nu\nu)$+jets and $W(\to \ell \nu)$+jets events, where for $W$+jets events the lepton is an unidentified electron or muon or
a hadronically decaying $\tau$-lepton. In the 1-lepton SRs, the subdominant backgrounds are
single-top, $\ttbar W$ and $\ttbar Z$.
 
\begin{figure}[htbp]
\centering
\subfigure[]{\includegraphics[width=0.9\textwidth]{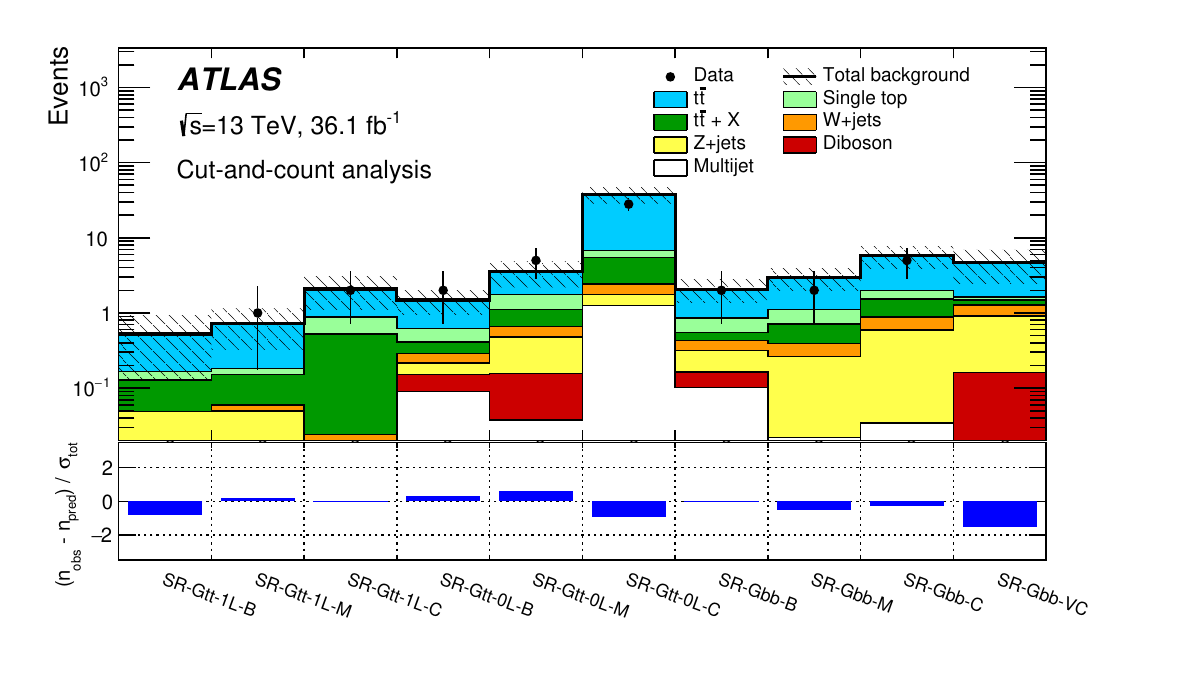}\label{fig:pullSR_discovery}}\\
\subfigure[]{\includegraphics[width=0.9\textwidth]{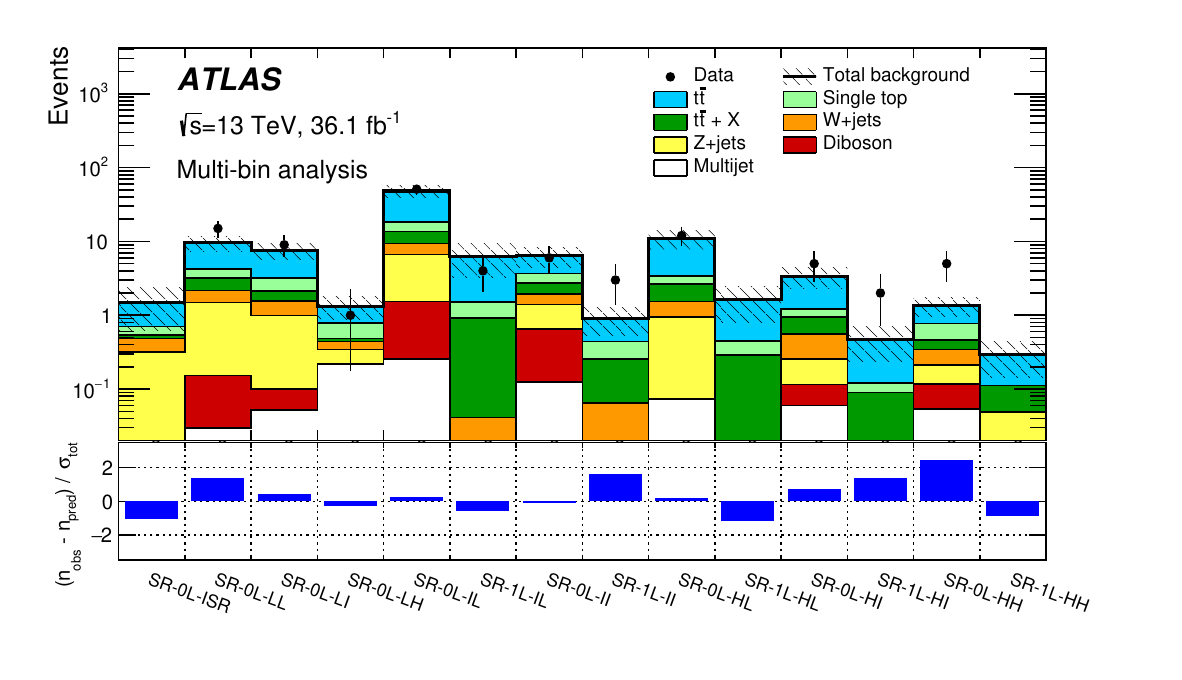}\label{fig:pullSR_exclusion}}\\
\caption{Results of the background-only fit extrapolated to the SRs for \subref{fig:pullSR_discovery}
the cut-and-count and \subref{fig:pullSR_exclusion} the multi-bin analyses. The data in the  SRs are
not included in the fit.  The upper panel shows the observed number of events and the predicted background
yield. All uncertainties  defined in Section~\ref{sec:syst} are included in the uncertainty band. The background
category $\ttbar+X$ includes $\ttbar W/Z$, $\ttbar H$ and $\ttbar \ttbar$ events. The lower panel shows the
pulls in each SR.}
\label{fig:pullSR}
\end{figure}

Table~\ref{tab:yield_discovery} shows the observed number of events and predicted number of background
events from the background-only fit in the Gtt 1-lepton, Gtt 0-lepton and Gbb regions for the cut-and-count analysis.
The central value of the fitted background is in general larger than the MC-only prediction. This is in part due to an
underestimation of the cross-section of $\ttbar + \geq \mathrm{1}b$ and $\ttbar + \geq \mathrm{1}c$ processes in
the simulation.
 
\begin{table*}[htbp]
\centering
\caption{Results of the background-only fit extrapolated to the Gtt 1-lepton, Gtt 0-lepton and Gbb SRs in
the cut-and-count analysis, for the total background prediction and breakdown of the main background sources.
The uncertainties shown include all systematic uncertainties. The data in the SRs are not included in the fit.
The background category $\ttbar+X$ includes $\ttbar W/Z$, $\ttbar H$ and $\ttbar \ttbar$ events.
The row ``MC-only background'' provides the total background prediction when the
$\ttbar$ normalisation is obtained from a theoretical
calculation~\cite{Czakon:2011xx}.
}
\label{tab:yield_discovery}
 
\begin{tabular}{lccc}
\toprule
& \multicolumn{3}{c}{SR-Gtt-1L} \\
\midrule
Targeted kinematics & B            &   M         &   C               \\[-0.05cm]
\midrule
Observed events              &  0   &  1 & 2  \\
\midrule
Fitted background             & 0.5 $\pm$ 0.4 & 0.7 $\pm$ 0.4 & 2.1 $\pm$ 1.0\\
\midrule
\ttbar\              &  0.4 $\pm$ 0.4 & 0.5 $\pm$ 0.4 & 1.2 $\pm$ 0.8\\
Single-top             & 0.04 $\pm$ 0.05 & 0.03 $\pm$ 0.06 & 0.35 $\pm$ 0.28\\
$\ttbar+X$          & 0.08 $\pm$ 0.05 & 0.09 $\pm$ 0.06 & 0.50 $\pm$ 0.28\\
$Z$+jets            & 0.049 $\pm$ 0.023 & 0.050 $\pm$ 0.023 & $<0.01$ \\
$W$+jets              & $<0.01$  & $<0.01$  & 0.024 $\pm$ 0.026\\
Diboson             & $<0.01$ & $<0.01$  & $<0.01$ \\
\midrule
MC-only background &  0.43 & 0.45 & 1.9 \\
\bottomrule
\end{tabular}
 
\vspace{0.4cm}
 
\begin{tabular}{lccc}
\toprule
& \multicolumn{3}{c}{SR-Gtt-0L} \\
\midrule
Targeted kinematics & B            &   M         &   C               \\[-0.05cm]
\midrule
Observed events              &  2 & 5 & 28 \\
\midrule
Fitted background             & 1.5 $\pm$ 0.5 & 3.5 $\pm$ 1.3 & 38 $\pm$ 8\phantom{0} \\
\midrule
\ttbar\              & 0.9 $\pm$ 0.4 & 1.8 $\pm$ 0.7 & 31 $\pm$ 8\phantom{0} \\
Single-top             & 0.21 $\pm$ 0.14 & 0.6 $\pm$ 0.4 & 1.3 $\pm$ 1.1\\
$\ttbar+X$          & 0.12 $\pm$ 0.07 & 0.45 $\pm$ 0.25 & 3.0 $\pm$ 1.6\\
$Z$+jets             & 0.06 $\pm$ 0.10 & 0.3 $\pm$ 0.9 & 0.49 $\pm$ 0.31\\
$W$+jets            & 0.07 $\pm$ 0.06 & 0.18 $\pm$ 0.15 & 0.67 $\pm$ 0.22\\
Diboson             &  0.06 $\pm$ 0.07 & 0.12 $\pm$ 0.07 & $<0.01$\\
Multijet               &  0.09 $\pm$ 0.11 & 0.04 $\pm$ 0.05 & 1.3 $\pm$ 2.1\\
\midrule
MC-only background &   1.3 & 3.3 & 23\\
\bottomrule
\end{tabular}
 
\vspace{0.4cm}
 
\begin{tabular}{lcccc}
\toprule
& \multicolumn{4}{c}{SR-Gbb} \\
\midrule
Targeted kinematics & B            	&   	M   		&   C   &   VC                \\[-0.05cm]
\midrule
Observed events              & 2 & 2 & 5 & 0\\
\midrule
Fitted background           &   2.1 $\pm$ 0.7 & 3.0 $\pm$ 1.0 & 5.8 $\pm$ 1.9 & 4.7 $\pm$ 2.3\\
\midrule
\ttbar\            			& 1.2 $\pm$ 0.6 & 1.9 $\pm$ 0.7 & 3.8 $\pm$ 1.3 & 3.1 $\pm$ 1.3\\
Single-top             		& 0.31 $\pm$ 0.16 & 0.39 $\pm$ 0.16 & 0.46 $\pm$ 0.20 & 0.15 $\pm$ 0.18\\
$\ttbar+X$             		& 0.12 $\pm$ 0.06 & 0.33 $\pm$ 0.19 & 0.6 $\pm$ 0.4 & 0.19 $\pm$ 0.11\\
$Z$+jets             		& 0.15 $\pm$ 0.34 & 0.2 $\pm$ 0.6 & 0.6 $\pm$ 1.3 & 0.8 $\pm$ 1.9\\
$W$+jets             		& 0.12 $\pm$ 0.09 & 0.13 $\pm$ 0.12 & 0.29 $\pm$ 0.19 & 0.37 $\pm$ 0.30\\
Diboson             		& 0.06 $\pm$ 0.04 & $<0.01$ & $<0.01$ & 0.15 $\pm$ 0.08\\
Multijet              & 0.10 $\pm$ 0.12 & 0.022 $\pm$ 0.025 & 0.03 $\pm$ 0.04 & 0.016 $\pm$ 0.020\\
\midrule
MC-only background & 1.9 & 2.7 & 4.4 & 3.9  \\
\bottomrule
\end{tabular}
\end{table*}
 
\section{Interpretation}
\label{sec:interpretation}
Since no significant excess over the expected background from SM processes is observed, the data are used to derive
one-sided upper limits at 95\% confidence level (CL). Two levels of interpretation are provided in this paper: model-independent exclusion
limits and model-dependent exclusion limits set on the Gbb, Gtt and gluino variable branching ratio models.
 
\subsection{Model-independent exclusion limits}
 
Model-independent limits on the number of beyond-the-SM (BSM) events for each SR are derived with
pseudoexperiments using the CL$_\mathrm{s}$ prescription \cite{Read:2002hq} and neglecting a possible signal contamination
in the CR. Only the single-bin regions from the cut-and-count analysis are used for this purpose, to aid in
the reintepretation of these limits. Limits are obtained with a fit in each SR which proceeds in the same way
as the fit used to predict the background, except that the number of events observed in the SR is added as an input to
the fit. Also, an additional parameter for the non-SM signal strength,
constrained to be non-negative, is fit.
Upper limits on the
visible BSM cross-section  ($\sigma^{95}_\mathrm{vis}$) are obtained
by dividing the observed upper limits on the number of BSM events with the
integrated luminosity.
The results are given in Table~\ref{mod-ind-lim}, where the $p_0$-values, which represent
the probability of the SM background alone to fluctuate to the observed number of events or higher, are also provided.
 
\begin{table}[t]
\centering
\caption{The $p_0$-values and $Z$ (the number of equivalent Gaussian standard deviations),
the 95$\%$ CL upper limits on the visible cross-section
($\sigma^{95}_\mathrm{vis}$),
and the observed and
expected 95$\%$ CL upper limits on the number of BSM events ($S_{\textrm
obs}^{95}$ and $S_{\textrm exp}^{95}$). The maximum
allowed $p_0$-value
is truncated at 0.5.}
\label{mod-ind-lim}
\small
\begin{tabular*}{0.6\textwidth}{@{\extracolsep{\fill}}lcccc}
\noalign{\smallskip}\toprule\noalign{\smallskip}
Signal channel         & $p_0$ (Z)            & $\sigma^{95}_\mathrm{vis}$ [fb]  &  $S_{\textrm obs}^{95}$  & $S_{\textrm exp}^{95}$   \\
\noalign{\smallskip}\midrule \noalign{\smallskip}
SR-Gtt-1L-B & $ 0.50~(0.00) $ &  $0.08$ &  $3.0$ & $ { 3.0 }^{ +1.0 }_{ -0.0 }$ \\[1mm]
SR-Gtt-1L-M & $ 0.34~(0.42)$ &  $0.11$ &  $3.9$ & $ { 3.6 }^{ +1.1 }_{ -0.4 }$ \\[1mm]
SR-Gtt-1L-C & $ 0.50~(0.00)$ &  $0.13$ &  $4.8$ & $ { 4.7 }^{ +1.8 }_{ -0.9 }$ \\[1mm]
\noalign{\smallskip}\midrule \noalign{\smallskip}
SR-Gtt-0L-B & $ 0.32~(0.48)$ & $0.13$ &  $4.8$ & $ { 4.1 }^{ +1.7 }_{ -0.6 }$  \\[1mm]
SR-Gtt-0L-M & $ 0.25~(0.69)$ &  $0.21$ &  $7.5$ & $ { 6.0 }^{ +2.3 }_{ -1.4 }$ \\[1mm]
SR-Gtt-0L-C & $ 0.50~(0.00)$ &  $0.39$ &  $14.0$ & $ { 17.8 }^{ +6.6 }_{ -4.5 }$ \\[1mm] 
\noalign{\smallskip}\midrule\noalign{\smallskip}
SR-Gbb-B & $ 0.50~(0.00) $ &  $0.13$ &  $4.6$ & $ { 4.6 }^{ +1.7 }_{ -1.0 }$  \\[1mm]
SR-Gbb-M & $ 0.50~(0.00) $ & $0.12$ &  $4.4$ & $ { 5.0 }^{ +1.9 }_{ -1.1 }$ \\[1mm]
SR-Gbb-C & $ 0.50~(0.00) $ &  $0.18$ &  $6.6$ & $ { 6.9 }^{ +2.7 }_{ -1.8 }$ \\[1mm]
SR-Gbb-VC & $ 0.50~(0.00) $ &  $0.08$ &  $3.0$ & $ { 4.6 }^{ +2.0 }_{ -1.3 }$\\
\noalign{\smallskip}\midrule\noalign{\smallskip}
\end{tabular*}
\end{table}
 
\subsection{Model-dependent exclusion limits}
 
The results are used to place exclusion limits on various signal models. The results are obtained using the
CL$_\mathrm{s}$ prescription
in the asymptotic approximation \cite{Cowan:2010js}. The expected and observed limits were compared to the CL$_\mathrm{s}$ calculated from pseudoexperiments and found to be compatible.
The signal contamination in the CRs
and the experimental systematic uncertainties in the signal are taken into account for this
calculation. All the regions of the multi-bin analysis are statistically combined to set model-dependent upper limits on the Gbb,
Gtt and variable branching ratio models.
 
The 95\%~CL observed and expected exclusion limits for the Gtt and Gbb models are shown
in the LSP and gluino mass plane in Figures~\ref{fig:limits_Gtt}  and~\ref{fig:limits_Gbb},
respectively. The $\pm1 \sigma^{\textrm SUSY}_{\textrm theory}$ lines around the observed limits
are obtained by changing the SUSY cross-section by one standard deviation ($\pm1\sigma$),
as described in Section~\ref{sec:samples}.  The yellow band around the expected limit shows
the $\pm1\sigma$ uncertainty, including all statistical and systematic uncertainties except the
theoretical uncertainties in the SUSY cross-section.
Compared to the previous results~\cite{Aad:2016eki}, the gluino mass sensitivities of the current search (assuming massless
LSPs) have improved by 300 GeV and 450 GeV for the Gbb and Gtt models,
respectively.
Gluinos with masses below 1.97~(1.92)~\tev\ are excluded at 95\%~CL for neutralino masses lower
than 300 GeV in the Gtt (Gbb) model. The observed limit for the Gtt
model at high gluino mass is weaker than the
expected limits due to the  mild excesses observed
in the signal regions SR-0L-HH and SR-1L-HI of the multi-bin fit analysis.
The best exclusion limit on the LSP mass is approximately 1.19
(1.20)~\tev, which is reached for a gluino mass of approximately 1.40 (1.68)~\tev\
for Gbb and Gtt models, respectively.
 
\begin{figure}[htbp]
\centering
\subfigure[]{\includegraphics[width=0.49\textwidth]{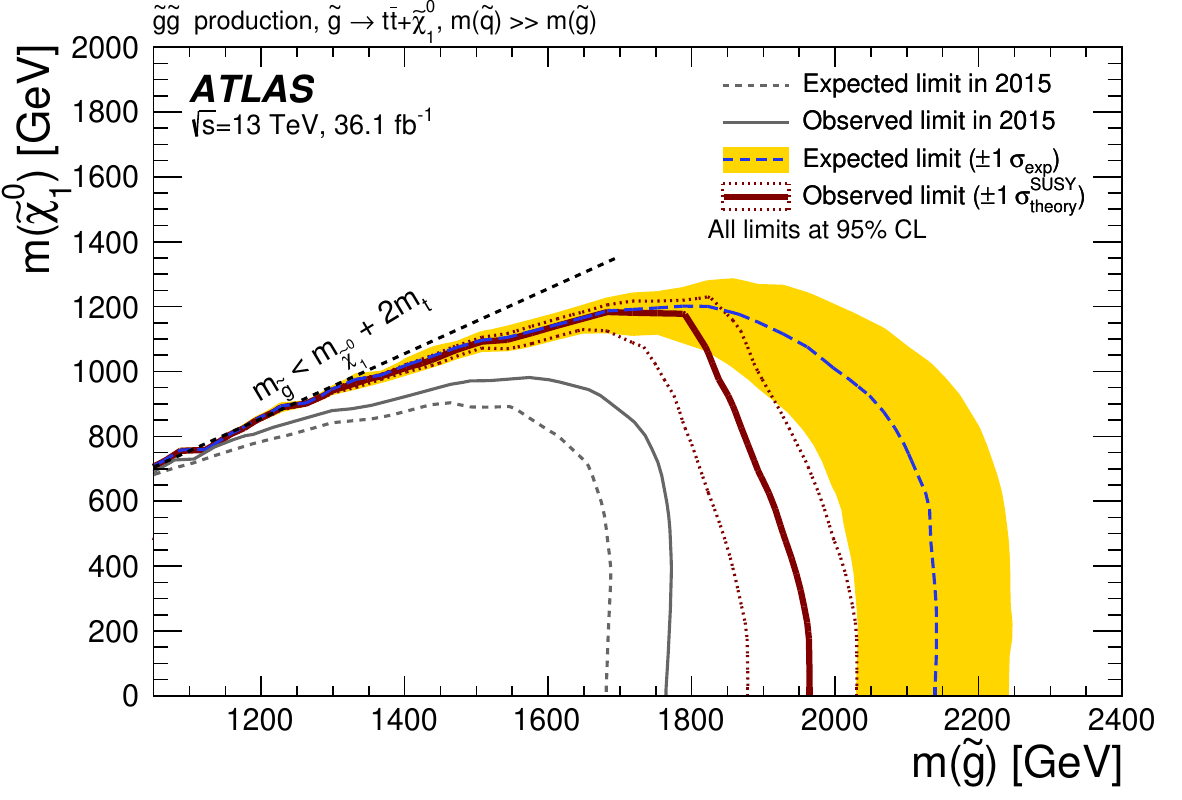}\label{fig:limits_Gtt}}
\subfigure[]{\includegraphics[width=0.49\textwidth]{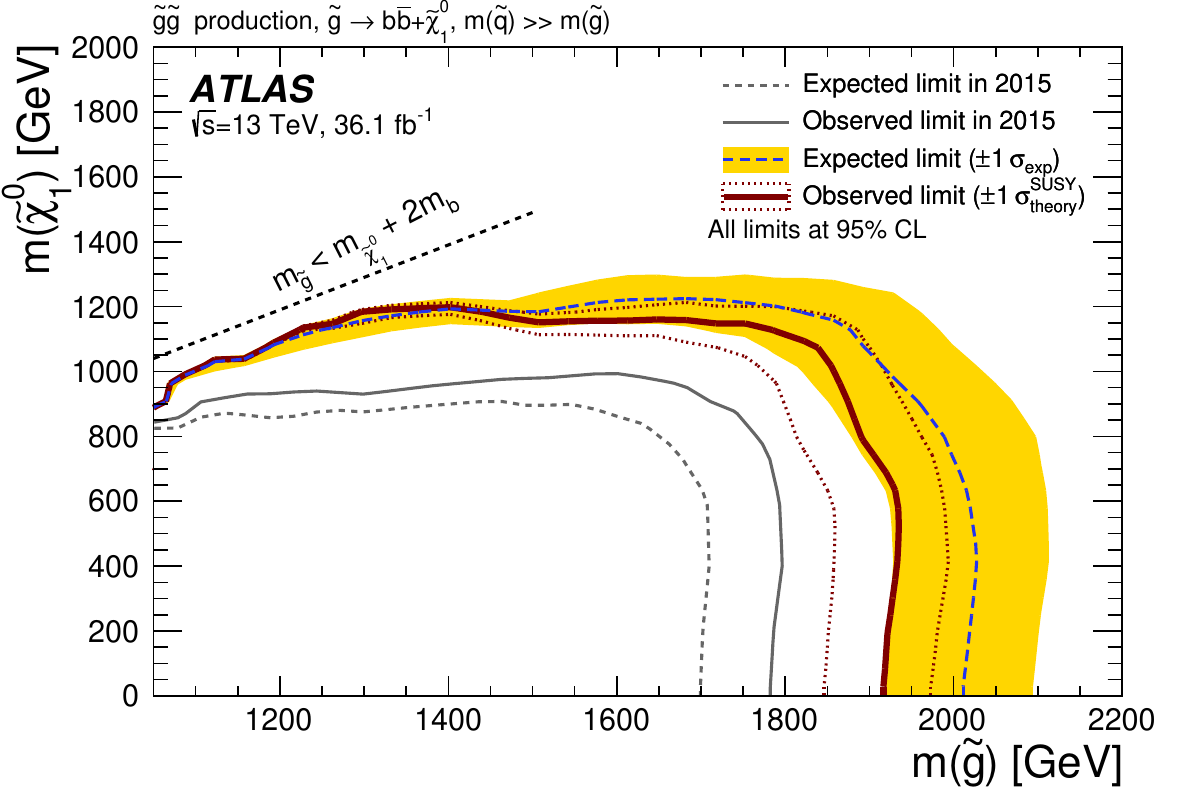}\label{fig:limits_Gbb}}
\caption{Exclusion limits in the $\ninoone$ and $\gluino$ mass plane
for the \subref{fig:limits_Gtt} Gtt and  \subref{fig:limits_Gbb} Gbb models obtained
in the context of the multi-bin analysis. The dashed and solid bold lines
show the 95\% CL expected and observed limits, respectively. The
shaded bands around the expected limits show the
impact of the
experimental and background uncertainties. The dotted
lines show the impact on the observed limit of the variation of the
nominal signal cross-section by $\pm 1 \sigma$ of its theoretical
uncertainty.
The 95\%~CL expected and observed limits from the ATLAS search based on 2015 data
\cite{Aad:2016eki} are also shown.}
\label{fig:limits_GbbGtt}
\end{figure}

Limits are also set in the signal model described in Section \ref{sec:susy_sig} for which the branching ratios
of the gluinos to $t \bar{b} \chinoonem$ (with $\chinoonem \to f\bar{f}' \ninoone)$, $ t \bar{t} \ninoone$, and $ b \bar{b} \ninoone$
are allowed to vary, with a unitarity constraint imposed on the sum of the three branching ratios. The expected and observed exclusions
are shown in Figure \ref{fig:limit_br_fixed_neu} for a fixed neutralino mass hypothesis ($m_{\ninoone} = 1$~\gev) and various gluino
masses. The results are presented in the $B(\gluino \to t \bar{t} \ninoone$) vs. $B(\gluino \to b \bar{b} \ninoone$) plane, where the
branching ratio for $\gluino \to t \bar{b} \chinoonem$ is equal to $1 - (B(\gluino \to t \bar{t} \ninoone) +B(\gluino \to b \bar{b} \ninoone))$.
The exclusion limits are weaker in the lower left corner, where the
branching ratio for $\gluino \to t \bar{b} \chinoonem$ is substantial, which is expected since
these decays were not included in the optimisation procedure.
Due to the mild excess observed in some regions of the multi-bin analysis and despite an expected sensitivity across the whole plane for a massless
neutralino hypothesis, the 95\% CL limit for a 1.8~\tev\ gluino is of $B(\gluino \to t \bar{t} \ninoone)\geq30$\%
($B(\gluino \to b\bar{b} \ninoone)\geq40$\%) when
assuming $B(\gluino \to b\bar{b} \ninoone) = 0$
($B(\gluino \to t \bar{t} \ninoone) =0$).
None of the points in the plane are excluded for gluino masses larger than 2.0~\tev.
 
Similar results are presented in Figure \ref{fig:limit_br_fixed_glu} assuming a gluino mass of 1.9~\tev\ and scanning
various neutralino masses (1, 600 and 1000~GeV). For neutralino masses between 1 and 600 \gev, most of the branching ratio plane
is expected to be excluded at 95\% CL. The observed limit is nevertheless worse due to the mild excess observed in the SRs.
Thus, for instance, for a massless neutralino hypothesis, only the region with
$B(\gluino \to b \bar{b} \ninoone)>90$~\% is excluded for all values of $B(\gluino \to t \bar{t} \ninoone)$.
 
\begin{figure}[htbp]
\centering
\subfigure[]{\includegraphics[width=0.49\textwidth]{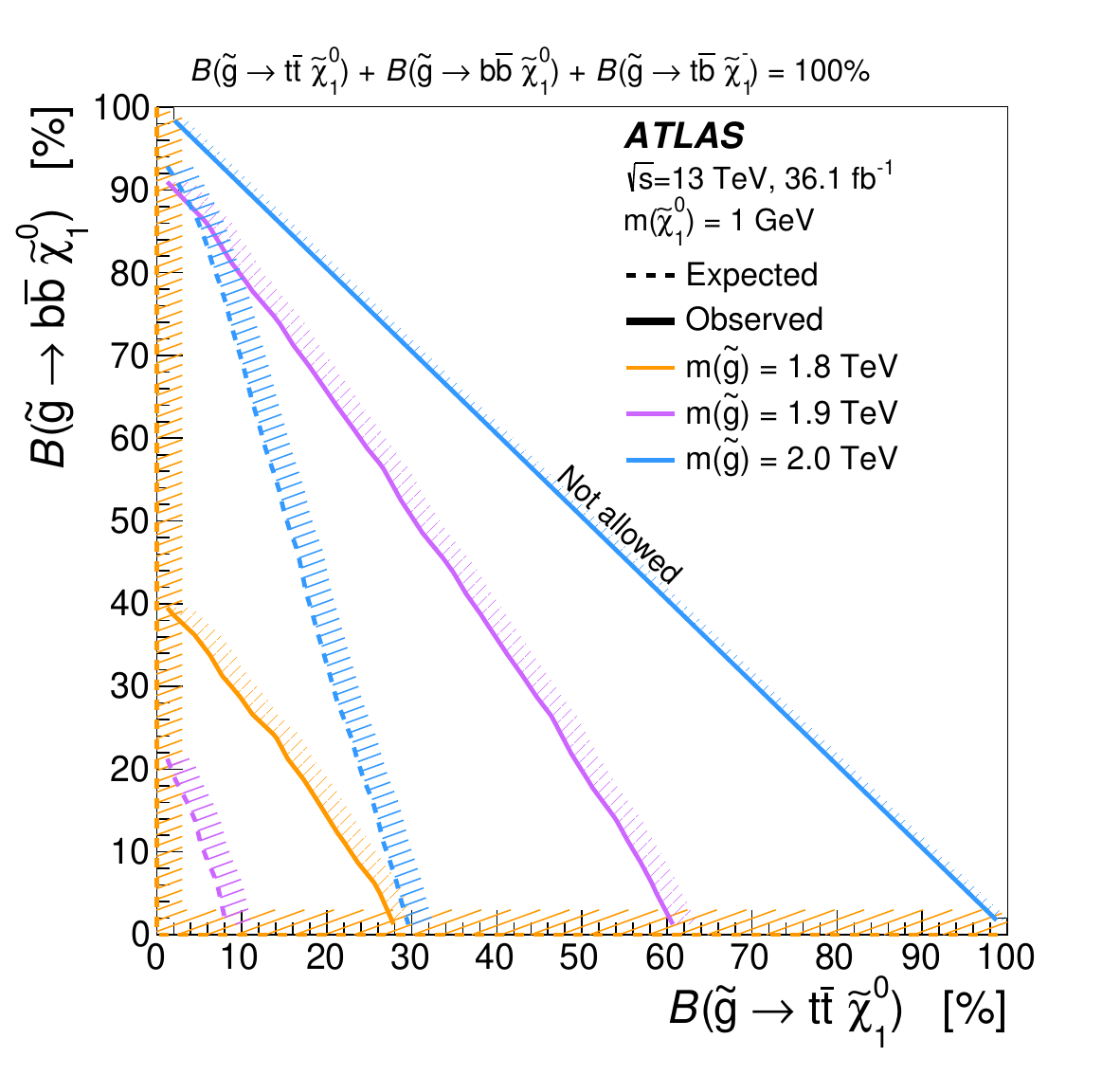}\label{fig:limit_br_fixed_neu}}
\subfigure[]{\includegraphics[width=0.49\textwidth]{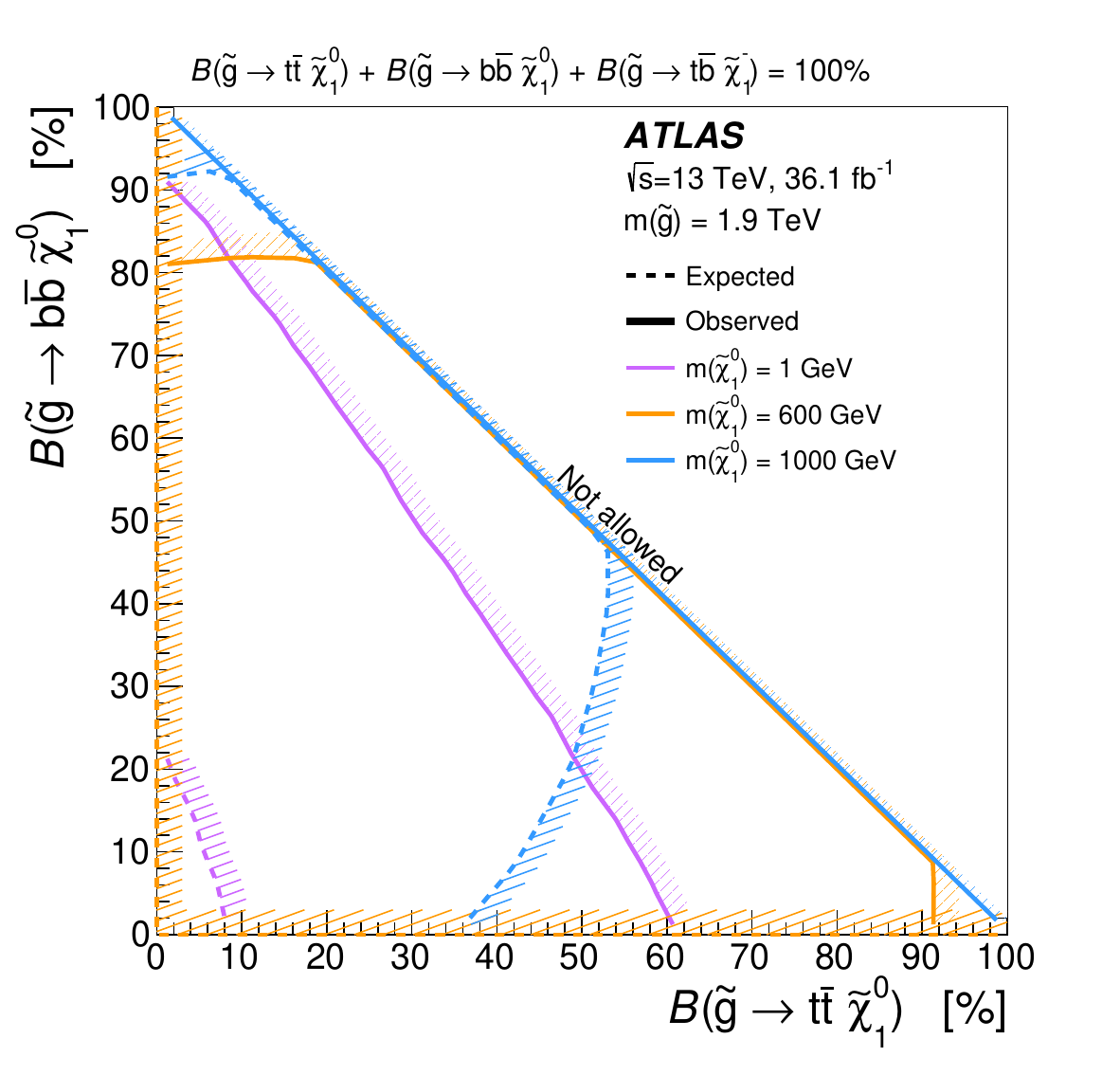}\label{fig:limit_br_fixed_glu}}
\caption{Exclusion limits in the $\gluino \to t \bar{t} \ninoone$ and $\gluino \to b \bar{b} \ninoone$
branching ratio plane assuming \subref{fig:limit_br_fixed_neu} a neutralino mass of 1~\gev\ and various gluino masses
(1.8, 1.9 and 2.0~\tev) and \subref{fig:limit_br_fixed_glu} a gluino mass of 1.9~\tev\ and three neutralino masses (1, 600 and 1000~\gev).
In \subref{fig:limit_br_fixed_neu}, the expected limit for a gluino mass of 1.8 \tev\ follows the plot axes, meaning that the whole plane is
expected to be excluded at 95\% CL. The same is true
in \subref{fig:limit_br_fixed_glu} for a neutralino
mass of 600~\gev.
The dashed and solid bold lines show the 95\% CL expected and observed limits, respectively. The hashing indicates which side of the line
is excluded. The upper right half of the plane is forbidden by the requirement that the sum of branching ratios does not exceed 100\%.}
\end{figure}

 
\FloatBarrier
\section{Conclusion}
\label{sec:conclusion}
A search for pair-produced gluinos decaying via bottom or top squarks is presented. LHC proton--proton
collision data from the full 2015 and 2016 data-taking periods are analysed, corresponding to an integrated
luminosity of  36.1~\ifb\ collected at $\rts =13\ \tev$ by the ATLAS detector.
The search uses multiple signal regions designed for different
scenarios of gluino and LSP masses. The signal regions require several high-$\pt$ jets,
of which at least three must be $b$-tagged, large $\met$ and either zero or at least one charged lepton.
Two strategies are employed: one in which the signal regions are optimised for discovery, and another one
in which several non-overlapping signal regions are fitted simultaneously to achieve optimal
exclusion limits for benchmark signals.
For all signal regions, the background is generally dominated by $\ttbar$+jets,
which is normalised in dedicated control regions.
No excess is found above the predicted background in any of the signal regions. Model-independent limits are
set on the visible cross-section for new physics processes. Exclusion limits are set on gluino and LSP
masses in two simplified models where the gluino decays exclusively as
$\gluino \to b\bar{b}\ninoone$ or $\gluino \to t\bar{t}\ninoone$.
For LSP masses
below approximately 300~\gev, gluino masses of less than 1.97~\tev\ and 1.92~\tev\ are excluded at the
95\% CL for the $\gluino \to t\bar{t}\ninoone$ and $\gluino \to b\bar{b}\ninoone$ models, respectively.
These results improve upon the exclusion limits obtained with
the 2015 dataset alone.
The results are also interpreted in a model with variable
gluino branching ratios to $\gluino \to b\bar{b}\ninoone$, $\gluino \to t \bar{b} \chinoonem$ and
$\gluino \to t\bar{t}\ninoone$. For example, a mass point with $m_{\gluino} = 1.9\ \tev$ and $m_{\ninoone} = 1~\gev$ is
excluded at the 95\% CL only if $B(\gluino \to t \bar{b} \chinoonem )$ < 10\%.
 
\section*{Acknowledgements}

We thank CERN for the very successful operation of the LHC, as well as the
support staff from our institutions without whom ATLAS could not be
operated efficiently.

We acknowledge the support of ANPCyT, Argentina; YerPhI, Armenia; ARC, Australia; BMWFW and FWF, Austria; ANAS, Azerbaijan; SSTC, Belarus; CNPq and FAPESP, Brazil; NSERC, NRC and CFI, Canada; CERN; CONICYT, Chile; CAS, MOST and NSFC, China; COLCIENCIAS, Colombia; MSMT CR, MPO CR and VSC CR, Czech Republic; DNRF and DNSRC, Denmark; IN2P3-CNRS, CEA-DRF/IRFU, France; SRNSFG, Georgia; BMBF, HGF, and MPG, Germany; GSRT, Greece; RGC, Hong Kong SAR, China; ISF, I-CORE and Benoziyo Center, Israel; INFN, Italy; MEXT and JSPS, Japan; CNRST, Morocco; NWO, Netherlands; RCN, Norway; MNiSW and NCN, Poland; FCT, Portugal; MNE/IFA, Romania; MES of Russia and NRC KI, Russian Federation; JINR; MESTD, Serbia; MSSR, Slovakia; ARRS and MIZ\v{S}, Slovenia; DST/NRF, South Africa; MINECO, Spain; SRC and Wallenberg Foundation, Sweden; SERI, SNSF and Cantons of Bern and Geneva, Switzerland; MOST, Taiwan; TAEK, Turkey; STFC, United Kingdom; DOE and NSF, United States of America. In addition, individual groups and members have received support from BCKDF, the Canada Council, CANARIE, CRC, Compute Canada, FQRNT, and the Ontario Innovation Trust, Canada; EPLANET, ERC, ERDF, FP7, Horizon 2020 and Marie Sk{\l}odowska-Curie Actions, European Union; Investissements d'Avenir Labex and Idex, ANR, R{\'e}gion Auvergne and Fondation Partager le Savoir, France; DFG and AvH Foundation, Germany; Herakleitos, Thales and Aristeia programmes co-financed by EU-ESF and the Greek NSRF; BSF, GIF and Minerva, Israel; BRF, Norway; CERCA Programme Generalitat de Catalunya, Generalitat Valenciana, Spain; the Royal Society and Leverhulme Trust, United Kingdom.

The crucial computing support from all WLCG partners is acknowledged gratefully, in particular from CERN, the ATLAS Tier-1 facilities at TRIUMF (Canada), NDGF (Denmark, Norway, Sweden), CC-IN2P3 (France), KIT/GridKA (Germany), INFN-CNAF (Italy), NL-T1 (Netherlands), PIC (Spain), ASGC (Taiwan), RAL (UK) and BNL (USA), the Tier-2 facilities worldwide and large non-WLCG resource providers. Major contributors of computing resources are listed in Ref.~\cite{ATL-GEN-PUB-2016-002}.

\printbibliography

\clearpage

\newpage
 
\begin{flushleft}
{\Large The ATLAS Collaboration}

\bigskip

M.~Aaboud$^\textrm{\scriptsize 34d}$,    
G.~Aad$^\textrm{\scriptsize 99}$,    
B.~Abbott$^\textrm{\scriptsize 124}$,    
O.~Abdinov$^\textrm{\scriptsize 13,*}$,    
B.~Abeloos$^\textrm{\scriptsize 128}$,    
S.H.~Abidi$^\textrm{\scriptsize 165}$,    
O.S.~AbouZeid$^\textrm{\scriptsize 143}$,    
N.L.~Abraham$^\textrm{\scriptsize 153}$,    
H.~Abramowicz$^\textrm{\scriptsize 159}$,    
H.~Abreu$^\textrm{\scriptsize 158}$
% ...
\bigskip
\\

% ...
$^{13}$Institute of Physics, Azerbaijan Academy of Sciences, Baku; Azerbaijan.\\
% ...
$^{34}$$^{(a)}$Facult\'e des Sciences Ain Chock, R\'eseau Universitaire de Physique des Hautes Energies - Universit\'e Hassan II, Casablanca;$^{(b)}$Centre National de l'Energie des Sciences Techniques Nucleaires (CNESTEN), Rabat;$^{(c)}$Facult\'e des Sciences Semlalia, Universit\'e Cadi Ayyad, LPHEA-Marrakech;$^{(d)}$Facult\'e des Sciences, Universit\'e Mohamed Premier and LPTPM, Oujda;$^{(e)}$Facult\'e des sciences, Universit\'e Mohammed V, Rabat; Morocco.\\
% ...
$^{99}$CPPM, Aix-Marseille Universit\'e, CNRS/IN2P3, Marseille; France.\\
% ...
$^{124}$Homer L. Dodge Department of Physics and Astronomy, University of Oklahoma, Norman OK; United States of America.\\
% ...
$^{128}$LAL, Universit\'e Paris-Sud, CNRS/IN2P3, Universit\'e Paris-Saclay, Orsay; France.\\
% ...
$^{143}$Santa Cruz Institute for Particle Physics, University of California Santa Cruz, Santa Cruz CA; United States of America.\\
% ...
$^{153}$Department of Physics and Astronomy, University of Sussex, Brighton; United Kingdom.\\
% ...
$^{158}$Department of Physics, Technion, Israel Institute of Technology, Haifa; Israel.\\
$^{159}$Raymond and Beverly Sackler School of Physics and Astronomy, Tel Aviv University, Tel Aviv; Israel.\\
% ...
$^{165}$Department of Physics, University of Toronto, Toronto ON; Canada.\\
% ...
$^{a}$ Also at  Department of Physics, University of Malaya, Kuala Lumpur; Malaysia.\\
$^{b}$ Also at Borough of Manhattan Community College, City University of New York, NY; United States of America.\\
$^{c}$ Also at Centre for High Performance Computing, CSIR Campus, Rosebank, Cape Town; South Africa.\\
$^{d}$ Also at CERN, Geneva; Switzerland.\\
$^{e}$ Also at CPPM, Aix-Marseille Universit\'e, CNRS/IN2P3, Marseille; France.\\
% ...
$^{*}$ Deceased

\end{flushleft}

 
\end{document}